\documentclass[12pt,oneside]{article}%
\usepackage{amsmath}
\usepackage{amssymb}
\usepackage{amsfonts}
\usepackage{graphicx}
\usepackage{cite}%
\allowdisplaybreaks
\newcommand{\eqa}{\begin{eqnarray}}
\newcommand{\ena}{\end{eqnarray}}

\def\one{\mbox{1 \kern-.59em {\rm l}}}
\begin{document}

\author{Alexander Schmidt\thanks{e-mail: schmidt@theorie.physik.uni-muenchen.de},
Hartmut Wachter\thanks{e-mail: Hartmut.Wachter@physik.uni-muenchen.de}
\\\hspace{0.4in}
\\Max-Planck-Institute\\for Mathematics in the Sciences
\\Inselstr. 22, D-04103 Leipzig, Germany
\\\hspace{0.4in}
\\Arnold-Sommerfeld-Center \\Ludwig-Maximilians-Universit\"{a}t\\Theresienstr. 37, D-80333 M\"{u}nchen, Germany}
\title{q-Deformed Superalgebras}
\date{}
\maketitle

\begin{abstract}
\noindent The article deals with $q$-analogs of the three- and
four-dimensional Euclidean superalgebra and the Poincar\'{e}
superalgebra.\newpage

\end{abstract}

\section{Introduction}

The idea of supersymmetry plays an important role in physics. Since its
invention \cite{Mya66, NS71, GS71, GL71, VA72, WZ74, Wessbagger} a great
experimental effort is undertaken to find supersymmetric particles in nature.
The reason for this lies in the fact that supersymmetry could provide
appealing solutions to outstanding problems in theoretical physics such as the
so-called hierarchy problem. Furthermore, there are indications that
supersymmetry can help in eliminating many of the divergences of certain
quantum field theories. If supersymmetry becomes a local gauge symmetry it
leads to supergravity, which is not as divergent as ordinary
gravity \cite{FNF76, DZ76}. Finally, the much-discussed superstring theories
propose the existence of space-time supersymmetry.

Let us recall that supersymmetry extends the symmetry algebra of space-time by
adding supersymmetry generators. In the last two decades, however, a much more
ambitious attempt to modify space-time symmetries has arisen. It is based on
noncommutative geometry \cite{Ku83, Wor87, RFT90, Tak90, Man88, CSSW90, PW90,
Lu92, Cas93, Dob94, DFR95, ChDe95, ChKu04, Koch04} and tries to modify
the whole space-time symmetry by deforming it in a consistent manner. There is
a great hope that such an approach yields a discretization of space-time
\cite{FLW96, CW98, MajReg, GKP96, Oec99, Wess00}\ which, in turn, implies an
effective method for regularizing quantum field theories \cite{Heis}.

In our previous work \cite{WW01, BW01, Wac02, Wac03, MSW04, SW04, qAn}\textbf{
}attention was concentrated on space-time structures that arise from
$q$-deformation \cite{WZ91}. More concretely, we are interested in
$q$-deformed quantum spaces that could prove useful in physical applications.
For this reason we deal with the three- and four-dimensional $q$-deformed
Euclidean space and the $q$-deformed Minkowski space \cite{Maj91, LWW97,
CSW91}. The symmetries of these quantum spaces are described by the
Drinfeld-Jimbo 
algebras $U_{q}(su(2))$ and $U_{q}(so(4))$ \cite{Dri85, Jim85, Drin86}
and the $q$-deformed Lorentz algebra \cite{SWZ91}. Finally, we can combine
these symmetry algebras with their quantum spaces and obtain $q$-analogs of
the three- and four-dimensional Euclidean algebra and the Poincar\'{e} algebra
\cite{OSWZ92, Maj93-2}.

It is an obvious thing to try to combine the ideas of supersymmetry with those
of deforming space-time symmetries (see, for example, Refs. \cite{FLM03,
Mik03, Man89, KT93, LA94, MR95, KLMS94, KLMS94-2, KLMS94-3, Witt98}) and the
aim of this article is to go the last step that makes the $q$-deformed
Poincar\'{e} algebra and the $q$-deformed Euclidean algebras in three and four
dimensions into superalgebras. To reach this goal we introduce supersymmetry
generators with spinor indices and make suitable ansaetze for their
commutation relations with the generators of the $q$-deformed Euclidean
algebras and the $q$-deformed Poincar\'{e} algebra. From consistency arguments
we get a system of equations for the unknown coefficients of our ansaetze.
Solving this system by a computer algebra system like Mathematica \cite{Wol}
we found that the relations of the three- and four-dimensional $q$-deformed
Euclidean superalgebra are uniquely determined and the same holds for the
$q$-deformed Poincar\'{e} superalgebra.

To write down our superalgebras in a rather compact form it is helpful to
recognize adjoint actions as $q$-analogs of classical commutators. Using
generators with definite transformation properties, these $q$-commutators can
often be expressed by $q$-deformed Pauli matrices and their relatives (for
their explicit form see Ref.\ \cite{qspinor1}). Last but not least, it should
be mentioned that in some sense the present article continues the reasonings
of Ref. \cite{qliealg}, where we adapted many general ideas about $q$-deformed
quantum algebras to our framework of conventions and notations.

\section{Symmetry algebras for three-di\-men\-sional $q$-deformed Euclidean
space}

In this section we first give a short review of the Hopf algebra
$U_{q}(su(2))$. Then, we combine this algebra with that of three-dimensional
$q$-deformed momentum space and obtain a $q$-analog of the three-dimensional
Euclidean algebra. These considerations culminate in the derivation of the
three-dimensional $q$-deformed Euclidean superalgebra.

\subsection{The Hopf algebra $U_{q}(su(2))$\label{uqsl2}}

The symmetry of $q$-deformed Euclidean space in three dimensions is described
by the quantum algebra $U_{q}(su(2))$ \cite{LWW97,Maj95, ChDe96, Klimyk}. This algebra is spanned by
the three generators $T^{+}$, $T^{-}$, and $T^{3}$ subject to the relations
\begin{align}
q^{-1}T^{+}T^{-}-qT^{-}T^{+}  &  =T^{3},\nonumber\\
q^{2}T^{3}T^{+}-q^{-2}T^{+}T^{3}  &  =\lambda_{+}T^{+},\nonumber\\
q^{-2}T^{3}T^{-}-q^{2}T^{-}T^{3}  &  =-\lambda_{+}T^{-}, \label{Talgebra}%
\end{align}
where $\lambda_{+}\equiv q+q^{-1}$. Instead of working with $T^{3}$, it is
often more convenient to use the generator $\tau^{3}$ defined by
\begin{equation}
\tau^{3}\equiv1-\lambda T^{3},\qquad\lambda\equiv q-q^{-1}.
\end{equation}
With this new generator the defining relations of $U_{q}(su(2))$ become
\begin{align}
T^{\pm}\tau^{3}  &  =q^{\pm4}\tau^{3}T^{\pm},\nonumber\\
T^{-}T^{+}  &  =q^{-2}T^{+}T^{-}+q^{-1}\lambda^{-1}(\tau^{3}-1).
\end{align}

It is well-known that the algebra $U_{q}(su(2))$ can be made into a Hopf
algebra. The corresponding coproduct, antipode, and counit on its generators
read as
\begin{align}
\Delta(T^{\pm})  &  =T^{\pm}\otimes1+(\tau^{3})^{1/2}\otimes T^{\pm
},\nonumber\\
\Delta(T^{3})  &  =T^{3}\otimes1+1\otimes T^{3}-\lambda T^{3}\otimes
T^{3},\nonumber\\
\Delta(\tau^{3})  &  =\tau^{3}\otimes\tau^{3},\\[0.08in]
S(T^{\pm})  &  =-(\tau^{3})^{-1/2}T^{\pm},\nonumber\\
S(T^{3})  &  =-(\tau^{3})^{-1}T^{3},\nonumber\\
S(\tau^{3})  &  =(\tau^{3})^{-1},\\[0.08in]
\epsilon(T^{A})  &  =0,\quad A\in{\{}+,3,-{\}},\nonumber\\
\epsilon(\tau^{3})  &  =1.
\end{align}

As symmetry algebra of three-dimensional q-deformed Euclidean space,
$U_{q}(su(2))$ can be viewed as $q$-analog of the algebra of three-dimensional
angular momentum. This becomes more clear if we rewrite the defining relations
of $U_{q}(su(2))$ by means of the new generators%
\begin{align}
L^{\pm}\equiv &  -q^{\pm1/2}\lambda_{+}^{-1/2}(\tau^{3})^{-1/2}T^{\pm
},\nonumber\\
L^{3}\equiv &  \,(\tau^{3})^{1/2}(L^{-}L^{+}-L^{+}L^{-})\nonumber\\
= &  \,-q\lambda^{-1}\lambda_{+}^{-1}(\tau^{3})^{-1/2}\big(q^{-2}\lambda^2T^{+}T^{-}%
+\tau^{3}-1\big).\label{L3T}%
\end{align}
In this manner, we get%
\begin{align}
L^{\pm}\tau^{3} &  =q^{\pm4}\tau^{3}L^{\pm},\qquad L^{3}\tau^{3}=\tau^{3}%
L^{3},\nonumber\\
L^{-}L^{+} &  =q^{2}L^{+}L^{-}+q\lambda^{-1}\lambda_{+}^{-1}((\tau^{3}%
)^{-1}-1),
\end{align}
or, alternatively,%
\begin{align}
L^{-}L^{+}-L^{+}L^{-} &  =(\tau^{3})^{-1/2}L^{3},\nonumber\\
L^{\pm}L^{3}-L^{3}L^{\pm} &  =q^{\pm1}L^{\pm}(\tau^{3})^{-1/2}%
.\label{VecUqsl2}%
\end{align}

In the classical limit $q\rightarrow1$ we regain the Liealgebra $su(2)$ from
the relations in (\ref{VecUqsl2}). This can easily be seen if we recognize
that $\tau^{3}$ tends to $1$ for $q\rightarrow1$. Thus, $L^{+}$, $L^{3}$, and
$L^{-}$ play the role of the components of $q$-deformed angular momentum in
three dimensions. Their coproducts, antipodes, and counits are found to be
\begin{align}
\Delta(L^{\pm})= &  \;L^{\pm}\otimes(\tau^{3})^{-1/2}+1\otimes L^{\pm
},\nonumber\\
\Delta(L^{3})= &  \;L^{3}\otimes(\tau^{3})^{-1/2}+(\tau^{3})^{1/2}\otimes
L^{3},\nonumber\\
&  +\lambda(\tau^{3})^{1/2}\big(q^{-1}L^{-}\otimes L^{+}+qL^{+}\otimes
L^{-}\big),\\[0.08in]
S(L^{\pm})= &  -L^{\pm}(\tau^{3})^{1/2},\nonumber\\
S(L^{3})= &  \;(\tau^{3})^{1/2}(q^{2}L^{+}L^{-}-q^{-2}L^{-}L^{+}),\\[0.08in]
\epsilon(L^{A})= &  \;0,\qquad A\in\{+,3,-\}.
\end{align}

In terms of the\ generators of $q$-deformed angular momentum, the Casimir
operator of $U_{q}(su(2))$ takes on the rather intuitive form \cite{qliealg}
\begin{equation}
L^{2}\equiv g_{AB}\,L^{A}L^{B}=-qL^{+}L^{-}+L^{3}L^{3}-q^{-1}L^{-}%
L^{+},\label{Lcaseu3}%
\end{equation}
where $g_{AB}$ denotes the quantum metric of three-dimensional $q$-deformed
Euclidean space. Notice that repeated indices are to be summed over if not
stated otherwise. The Casimir property of $L^2$ can be shown most
easily by making 
use of the relations (\ref{VecUqsl2}).

It should also be noted that the components of three-dimensional angular
momentum give rise to a quantum Lie algebra. To this end we introduce
$q$-analogs of classical commutators. These so-called $q$-commutators are
determined by the Hopf structure of $U_{q}(su(2))$ \cite{qliealg, Schupp93,
Sud92, Bie90, Maj94}:
\begin{equation}
{[}L^{A},L^{B}{]}_{q}\equiv L_{(1)}^{A}\,L^{B}S(L_{(2)}^{A})=S^{-1}%
(L_{(2)}^{B})L^{A}L_{(1)}^{B},\label{q-KommAllg}%
\end{equation}
where we have written the coproduct in Sweedler notation. It is now
straightforward\ to check that the relations in (\ref{VecUqsl2}) are
equivalent to (see, for example Ref. \cite{qliealg}):%
\begin{equation}
{[}L^{A},L^{B}{]}_{q}=q^{2}\varepsilon^{ABC}g_{CD}\,L^{D},
\end{equation}
where $\varepsilon^{ABC}$ denotes the three-dimensional $q$-deformed epsilon
tensor (its non-vanishing components are listed in the appendix).

The components $L^{+}$, $L^{3}$, and $L^{-}$ transform under $U_{q}(su(2))$ as
a vector. Using the three-dimensional $q$-deformed epsilon tensor we are able
to assign the components of angular momentum an antisymmetric tensor $M^{AB}$
\cite{qliealg}:%
\begin{equation}
M^{AB}\equiv k_{1}\,\varepsilon^{ABC}g_{CD}\,L^{D}, \quad k_1\in\mathbb{R}.\label{Mgeneu3}%
\end{equation}
More explicitly, we have as independent generators%
\begin{gather}
M^{+3}=-k_{1}q^{-1}L^{+},\quad M^{3-}=-k_{1}q^{-1}L^{-},\nonumber\\
M^{+-}=-k_{1}q^{-2}L^{3},\label{Mgeneu32}%
\end{gather}
with the additional conditions
\begin{gather}
M^{+-}=-M^{-+}, \quad M^{33}=\lambda M^{+-}, \nonumber\\
M^{++}=M^{--}=0,\quad
M^{\pm 3}=-q^{\mp 2}M^{3\pm}. 
\end{gather}
We regain the components of three-dimensional angular momentum through
\begin{equation}
L^{D}=k_{1}^{\prime}\,g^{DC}\,\varepsilon_{BAC}\,M^{AB},\quad k_1^{\prime}\in\mathbb{R},\label{Mgeneu3inv}%
\end{equation}
being tantamount to%
\begin{align}
L^{+} &  =-k_{1}^{\prime}q^{-3}(q^{2}+q^{-2})M^{+3},\nonumber\\
L^{3} &  =-k_{1}^{\prime}q^{-2}(q^{2}+q^{-2})M^{+-},\nonumber\\
L^{-} &  =-k_{1}^{\prime}q^{-3}(q^{2}+q^{-2})M^{3-},
\end{align}
where our choice of conventions requires that%
\begin{equation}
k_{1}k_{1}^{\prime}=q^{4}(q^{2}+q^{-2})^{-1}.
\end{equation}

Using the generators in (\ref{Mgeneu3}) the Casimir in (\ref{Lcaseu3}) reads
as%
\begin{equation}
L^{2}=(k_{1}^{\prime})^{2}q^{-4}(q^{2}+q^{-2})\,g_{AD}\,g_{BC}%
\,M^{AB}\,M^{CD}.
\end{equation}
In the same manner, the relations for the quantum Lie algebra of
$U_{q}(su(2))$ turn into 
\begin{equation}
{[}M^{AB},M^{CD}{]}_{q}=(k_1^{\prime}{})^{-1}\frac{q^{2}(q^{2}+q^{-2})}%
{q^{2}-1+q^{-2}}(P_{A})^{AB}{}_{EF}(P_{A})^{CD}{}_{GH}\,g^{FG}M^{EH},
\end{equation}
where $P_{A}$ stands for the antisymmetrizer of three-dimensional $q$-deformed
Euclidean space \cite{LWW97}. More explicitly,
\begin{align}
{[}M^{+3},M^{+-}{]}_{q} &  =-q^{2}{[}M^{+-},M^{+3}{]}_{q}=(k_1^{\prime}{})%
^{-1}q^{3}(q^{2}+q^{-2})^{-1}M^{+3},\nonumber\\
{[}M^{+3},M^{3-}{]}_{q} &  =-{[}M^{3-},M^{+3}{]}_{q}=(k_1^{\prime}{})^{-1}%
q^{4}(q^{2}+q^{-2})^{-1}M^{+-},\nonumber\\
{[}M^{+-},M^{3-}{]}_{q} &  =-q^{2}{[}M^{3-},M^{+-}{]}_{q}=(k_1^{\prime}{})%
^{-1}q^{3}(q^{2}+q^{-2})^{-1}M^{3-},\nonumber\\
{[}M^{+-},M^{+-}{]}_{q} &
=(k_1^{\prime}{})^{-1}q^2(q^2+q^{-2})^{-1}\lambda M^{+-}.
\end{align}
Notice that in analogy to (\ref{q-KommAllg}) we have%
\begin{equation}
{[}M^{AB},M^{CD}{]}_{q}\equiv M_{(1)}^{AB}M^{CD}S(M_{(2)}^{AB})=S^{-1}%
(M_{(2)}^{CD})M^{AB}M_{(1)}^{CD}.\label{qKomTens}%
\end{equation}

\subsection{The three-dimensional $q$-deformed Euclidean algebra}

Next, we wish to combine the algebra of three-dimensional angular momentum
with the algebra of $q$-deformed momentum space. This way, we arrive at a
$q$-analog of the three-dimensional $q$-deformed Euclidean algebra. First of
all, let us recall the commutation relations for the three-dimensional
momentum generators $P^{A}$, $A\in\{+,3,-\}$:%
\begin{equation}
(P_{A})^{AB}{}_{CD}\,P^{C}P^{D}=0.
\end{equation}
The above condition implies as independent
relations%
\begin{equation}
P^{3}P^{\pm}-q^{\pm2}P^{\pm}P^{3}=0,\qquad P^{-}P^{+}-P^{+}P^{-}=\lambda
P^{3}P^{3}.
\end{equation}

It remains to specify the commutation relations between the generators of
$U_{q}(su(2))$ and the momentum algebra. To this end, we first have to realize
that the momentum generators establish a vector representation of
$U_{q}(su(2))$. The point now is that we can combine a Hopf algebra
$\mathcal{H}$ with its representation space $\mathcal{A}$ to form a so-called
left-cross product algebra $\mathcal{A}\rtimes\mathcal{H}$ \cite{Maj95,
Klimyk, ChDe96} built on $\mathcal{A}\otimes\mathcal{H}$ with product%
\begin{equation}
(a\otimes h)(b\otimes g)=a(h_{(1)}\triangleright b)\otimes h_{(2)}\,g,\quad
a,b\in\mathcal{A},\mathcal{\quad}h,g\in\mathcal{H}.\label{LefCrosPro}%
\end{equation}
The last identity tells us that the commutation relations between symmetry
generators and representation space elements are completely determined by
coproduct and action of the Hopf algebra $\mathcal{H}$, since we have%
\begin{equation}
hb=(1\otimes h)(b\otimes1)=(h_{(1)}\triangleright b)\otimes h_{(2)}.
\end{equation}

Applying these ideas to $U_{q}(su(2))$ and the three-dimensional $q$-deformed
momentum algebra we obtain the relations%
\begin{gather}
L^{\pm}P^{\pm}-P^{\pm}L^{\pm}=0,\nonumber\\
L^{\pm}P^{\mp}-P^{\mp}L^{\pm}=\mp P^{3}\tau^{-1/2},\nonumber\\
L^{\pm}P^{3}-P^{3}L^{\pm}=\mp q^{\pm1}P^{\pm}\tau^{-1/2},\nonumber\\
L^{3}P^{\pm}-q^{\mp2}P^{\pm}L^{3}=\pm q^{\mp1}\lambda P^{3}L^{\pm}\pm q^{\mp
1}P^{\pm}\tau^{-1/2},\nonumber\\
L^{3}P^{3}-P^{3}L^{3}=\lambda(P^{-}L^{+}-P^{+}L^{-})-\lambda P^{3}\tau
^{-1/2}.\label{ComLP}%
\end{gather}
By means of the $q$-commutator%
\begin{equation}
{[}L^{A},P^{B}{]}_{q}\equiv L^{A}\triangleright P^{B}=L_{(1)}^{A}%
\,P^{B}S(L_{(2)}^{A}),
\end{equation}
which is nothing other than the adjoint action of $L^{A}$ on $P^{B}$, the
relations in (\ref{ComLP}) can be written more compactly:%
\begin{equation}
{[}L^{A},P^{B}{]}_{q}=q^{2}\,\varepsilon^{ABC}g_{CD}\,P^{D}.
\end{equation}
Using (\ref{Mgeneu3inv}) one can verify that
\begin{equation}
{[}M^{AB},P^C{]}_{q}=(k_1^{\prime})^{-1}q^2(P_A)^{AB}{}_{B^{\prime}C^{\prime}}\eta^{C^{\prime}C}P^{B^{\prime}}.
\end{equation}

Last but not least, we would like to give the Casimir operators of the
three-dimensional $q$-deformed Euclidean algebra. With the commutation
relations presented so far one can verify that the operators
\begin{equation}
C_{1}\equiv g_{AB}\,P^{A}P^{B},\qquad C_{2}\equiv g_{AB}\,L^{A}P^{B}%
\end{equation}
commute with all generators of $U_{q}(su(2))$ and the momentum algebra.

\subsection{Symmetry algebra in spinor notation}

In Ref. \cite{qspinor1} we discussed $q$-analogs of Pauli matrices that enable
us to construct a vector out of two spinors. On these grounds, we can use
$q$-deformed\ Pauli matrices to switch from the vectorial generators $L^{A}$
to operators $M^{\alpha\beta}$ with two spinor indices:%
\begin{equation}
M^{\alpha\beta}=k_{2}^{\prime}\,(\sigma_{A}^{-1})^{\alpha\beta}L^{A},
\quad k_2^{\prime}\in \mathbb{R}.
\end{equation}
From the completeness relation
\begin{equation}
(\sigma^{A})_{\alpha\beta}(\sigma_{B}^{-1})^{\alpha\beta}=\delta_{B}^{A},
\end{equation}
we get, at once,
\begin{equation}
L^{A}=k_{2}\,(\sigma^{A})_{\alpha\beta}\,M^{\alpha\beta}, \quad k_2
\in \mathbb{R}\label{Mspineu3}%
\end{equation}
with $k_{2}k_{2}^{\prime}=1$. Explicitly, we have%
\begin{gather}
M^{11}=k_{2}^{\prime}\,q^{-1}L^{-},\qquad M^{22}=k_{2}^{\prime}\,q^{-1}%
L^{+},\nonumber\\
M^{12}=k_{2}^{\prime}\,q^{-1/2}\lambda_{+}^{-1/2}L^{3},\qquad M^{21}%
=k_{2}^{\prime}\,q^{-3/2}\lambda_{+}^{-1/2}L^{3},\label{Mspineu32}%
\end{gather}
and
\begin{align}
L^{+} &  =k_{2}\,qM^{22},\qquad L^{-}=k_{2}\,qM^{11},\nonumber\\
L^{3} &  =k_{2}\,q\lambda_{+}^{-1/2}(q^{1/2}M^{12}+q^{-1/2}M^{21})\nonumber\\
&  =k_{2}\,q\,^{1/2}\lambda_{+}^{1/2}M^{12}.\label{Mspineu31}%
\end{align}

The matrix entries of the Pauli matrices $\sigma^{A}$ and $\sigma_{A}^{-1}$
can be looked up in Ref. \cite{qspinor1}. It should also be mentioned that the
so-called 'inverse' Pauli matrices $\sigma_{A}^{-1}$ should not be confused
with matrices being inverse in the sense of matrix multiplication (for the
details see again Ref. \cite{qspinor1}). Especially, we have the
identification
\begin{equation}
(\sigma_{A}^{-1})^{\alpha\beta}=q^{-2}(\sigma^{A})_{\alpha\beta},\qquad
A=\{+,3,-\}.
\end{equation}

We can also start our considerations from the generators $M^{AB}$ introduced
in (\ref{Mgeneu3}). They are related to the generators $M^{\alpha\beta}$ by
the formulae%
\begin{equation}
M^{\alpha\beta}=k_{3}^{\prime}\,(\sigma_{AB}^{-1})^{\alpha\beta}M^{AB},\qquad
M^{AB}=k_{3}\,(\sigma^{AB})_{\alpha\beta}M^{\alpha\beta},\label{ConvForm}%
\end{equation}
where $\sigma^{AB}$ and $\sigma_{AB}^{-1}$ denote the two-dimensional spin
matrices of three-dimensional $q$-deformed Euclidean space. For their explicit
form we refer the reader again to Ref. \cite{qspinor1}. Written out
explicitly, the relations in (\ref{ConvForm}) become
\begin{gather}
M^{11}=-k_{3}^{\prime}\,q^{-5}(q^4+1)\lambda_{+}^{-1/2}M^{3-},\qquad M^{22}%
=-k_{3}^{\prime}\,q^{-5}(q^4+1)\lambda_{+}^{-1/2}M^{+3},\nonumber\\
M^{12}=qM^{21}=-k_{3}^{\prime}\,q^{-7/2}(q^4+1)\lambda_{+}^{-1}M^{+-},
\end{gather}
and
\begin{gather}
M^{+3}=k_{3}\,q^{3}\lambda_{+}^{-1/2}M^{22},\qquad M^{3-}=k_{3}\,q^{3}%
\lambda_{+}^{-1/2}M^{11},\nonumber\\
M^{+-}=\lambda M^{33}=k_{3}\,q^{3/2}M^{12}.\label{Mkopp}%
\end{gather}

The two-dimensional spin matrices are subject to the completeness relation
\begin{equation}
(\sigma^{AB})_{\alpha\beta}(\sigma_{CD}^{-1})^{\alpha\beta}=-q^{-2}%
(q^{4}+1)\lambda_{+}^{-1}(P_{A})^{AB}{}_{CD}.
\end{equation}
This relation implies
\begin{equation}
k_{3}k_{3}^{\prime}=-q^{2}(q^{4}+1)^{-1}\lambda_{+},
\end{equation}
as can be proven by inserting the two equations of (\ref{ConvForm}) into each
other. Finally, it should be mentioned that the relations in (\ref{ConvForm})
are consistent with (\ref{Mgeneu3}) and (\ref{Mgeneu3inv}) iff the following
condition is satisfied:%
\begin{equation}
k_{1}k_{2}=-q^{3}\lambda_{+}^{-1/2}\,k_{3}.
\end{equation}

Last but not least, we use the generators $M^{\alpha\beta}$ with two spinor
indices to write the quantum Lie algebra and the Casimir operator of
$U_{q}(su(2))$ in an alternative form. Clearly, the $q$-commutators between
the $M^{\alpha\beta}$ are defined in complete analogy to (\ref{qKomTens}).
Using the correspondence between the $M^{\alpha\beta}$ and the vectorial
generators $L^{A}$ one can show that
\begin{align}
{[}M^{12},M^{12}{]}_{q} &  =-k_{2}^{-1}\,q^{-1/2}\lambda\lambda_{+}%
^{-1/2}M^{12},\nonumber\\
{[}M^{11},M^{12}{]}_{q} &  =-q^{-2}{[}M^{12},M^{11}{]}_{q}=k_{2}%
^{-1}\,q^{-3/2}\lambda_{+}^{-1/2}M^{11},\nonumber\\
{[}M^{11},M^{22}{]}_{q} &  =-{[}M^{22},M^{11}{]}_{q}=k_{2}^{-1}\,q^{-3/2}%
\lambda_{+}^{1/2}M^{12},\nonumber\\
{[}M^{12},M^{22}{]}_{q} &  =-q^{-2}{[}M^{22},M^{12}{]}_{q}=k_{2}%
^{-1}\,q^{-3/2}\lambda_{+}^{-1/2}M^{22},\label{algspineu3}%
\end{align}
and the remaining $q$-commutators all vanish. Inserting (\ref{Mspineu31}) in
(\ref{Lcaseu3}) we find the following expression for the Casimir operator of
$U_{q}(su(2))$:%
\begin{equation}
C=-k_{2}^{2}(q^{3}M^{11}M^{22}+qM^{22}M^{11}-q\lambda_{+}M^{12}M^{12}%
).\label{Casimirspineu3}%
\end{equation}

The $q$-commutators in (\ref{algspineu3}) and the Casimir in
(\ref{Casimirspineu3}) can be written in a closed form by using the
$q$-deformed spinor metric $\varepsilon_{\alpha\beta}$ and the symmetric
projector $S$ corresponding to the $\hat{R}$-matrix of $U_{q}(su(2))$. In this
manner, we have
\begin{align}
\lbrack M^{\alpha\beta},M^{\gamma\delta}]_{q} &  =-k_{2}^{-1}\,q^{-1}%
\lambda_{+}^{1/2}S^{\alpha\beta}{}_{\beta^{\prime}\alpha^{\prime}}%
S^{\gamma\delta}{}_{\delta^{\prime}\gamma^{\prime}}\,\varepsilon
^{\alpha^{\prime}\delta^{\prime}}M^{\beta^{\prime}\gamma^{\prime}},\nonumber\\
C &  =-k_{2}^{2}\,q^{2}\,\varepsilon_{\alpha\beta}\varepsilon_{\alpha^{\prime
}\beta^{\prime}}\,M^{\alpha^{\prime}\alpha}M^{\beta\beta^{\prime}}.
\end{align}

\subsection{The three-dimensional $q$-deformed Euclidean
superalgebra\label{SecSupAlg3dim}}

In the previous sections we considered the $q$-deformed Euclidean algebra in
three dimensions. This algebra is a cross product of $U_{q}(su(2))$ and
three-dimensional $q$-deformed Euclidean space. In this sense, it is spanned
by the generators of $U_{q}(su(2))$ and the components of three-dimensional
$q$-deformed momentum.

For implementing supersymmetry on $q$-deformed Euclidean space in three
dimensions, one has to extend its Euclidean algebra to a $q$-deformed
Euclidean superalgebra. To describe a $q$-deformed version of $N=1$
supersymmetry we have to add supersymmetry generators $Q^{\alpha}$ and
$\bar{Q}^{\alpha}$ with spinor indices. It is now our task to determine the
commutation relations concerning the new generators.

We assume that $Q^{1}$ together with $Q^{2}$ generate an antisymmetrized
quantum plane and the same should hold for $\bar{Q}^{1}$ and $\bar{Q}^{2}$.
Thus, we have%
\begin{align}
Q^{\alpha}Q^{\beta} &  =-q\hat{R}^{\alpha\beta}{}_{\alpha^{\prime}%
\beta^{\prime}}Q^{\alpha^{\prime}}Q^{\beta^{\prime}},\nonumber\\
\bar{Q}^{\alpha}\bar{Q}^{\beta} &  =-q\hat{R}^{\alpha\beta}{}_{\alpha^{\prime
}\beta^{\prime}}\bar{Q}^{\alpha^{\prime}}\bar{Q}^{\beta^{\prime}%
},\label{QQ-Rel}%
\end{align}
where $\hat{R}$ stands for the $\hat{R}$-matrix of $U_{q}(su(2))$. From
(\ref{QQ-Rel}) we find as independent relations%
\begin{gather}
Q^{\alpha}Q^{\alpha}=\bar{Q}^{\dot{\alpha}}\bar{Q}^{\dot{\alpha}%
}=0,\nonumber\\[0.03in]
Q^{1}Q^{2}=-q^{-1}Q^{2}Q^{1},\qquad\bar{Q}^{1}\bar{Q}^{2}=-q^{-1}\bar{Q}%
^{2}\bar{Q}^{1}.\label{qplanerel}%
\end{gather}
If we introduce a $q$-deformed anticommutator for spinor operators by%
\begin{equation}
{\{}\theta^{\alpha},\tilde{\theta}^{\beta}{\}}_{k}\equiv{\theta}%
^{\alpha}\tilde\theta^{\beta}+k\hat{R}^{\alpha\beta}{}_{\alpha^{\prime}\beta
^{\prime}}\,\tilde\theta^{\alpha^{\prime}}{\theta}^{\beta^{\prime}},
\end{equation}
the relations in (\ref{QQ-Rel}) become
\begin{equation}
\{Q^{\alpha},Q^{\beta}\}_{q}=\{\bar{Q}^{\alpha},\bar{Q}^{\beta}\}_{q}=0.
\end{equation}

It is obvious that $Q^{\alpha}$ as well as $\bar{Q}^{\alpha}$ establish
spin-$1/2$ representations of $U_{q}(su(2))$. This observation fixes the
commutation relations of the supersymmetry generators with the generators of
$U_{q}(su(2))$. As it was shown in Ref. \cite{qliealg} the spinor operators
$Q^{\alpha}$ and $\bar{Q}^{\alpha}$ then have to fulfill%
\begin{align}
{[}L^{A},Q^{\alpha}{]}_{q} &  =-q^{-1}\lambda_{+}^{-1/2}(\sigma^{A})_{\beta
}{}^{\alpha}Q^{\beta},\nonumber\\
{[}L^{A},\bar{Q}^{\alpha}{]}_{q} &  =-q^{-1}\lambda_{+}^{-1/2}(\sigma
^{A})_{\beta}{}^{\alpha}\bar{Q}^{\beta},\label{TransTensOp}%
\end{align}
where the $q$-commutator is a kind of shorthand notation for the adjoint
action, i.e.%
\begin{equation}
{[}L^{A},V{]}_{q}\equiv L_{(1)}^{A}VS(L_{(2)}^{A}).
\end{equation}
Notice that in (\ref{TransTensOp}) we used the $q$-deformed Pauli matrices $\sigma^{A}$
given in Ref. \cite{qspinor1}.\ If we instead work with the generators
(\ref{Mgeneu32}) or (\ref{Mspineu32}) we respectively have
\begin{align}
\lbrack M^{AB},Q^{\alpha}]_{q} &  =(k_{1}^{\prime})^{-1}(q^{2}%
+q^{-2})^{-1}\,Q^{\beta}(\sigma^{AB})_{\beta}{}^{\alpha},\nonumber\\
\lbrack M^{AB},\bar{Q}^{\alpha}]_{q} &  =(k_{1}^{\prime})^{-1}%
(q^{2}+q^{-2})^{-1}\,\bar{Q}^{\beta}(\sigma^{AB})_{\beta}{}^{\alpha},
\end{align}
and%
\begin{align}
\lbrack M^{\alpha\beta},Q^{\gamma}]_{q} &  =-k_{2}^{-1}q^{-1}\lambda
_{+}^{-1/2}S^{\alpha\beta}{}_{\alpha^{\prime}\beta^{\prime}}\,\varepsilon
^{\beta^{\prime}\gamma}Q^{\alpha^{\prime}},\nonumber\\
\lbrack M^{\alpha\beta},\bar{Q}^{\gamma}]_{q} &  =-k_{2}^{-1}q^{-1}\lambda
_{+}^{-1/2}S^{\alpha\beta}{}_{\alpha^{\prime}\beta^{\prime}}\,\varepsilon
^{\beta^{\prime}\gamma}\bar{Q}^{\alpha^{\prime}}.
\end{align}

It remains to find the commutation relations between $Q^{\alpha}$ and $\bar
{Q}^{\alpha}$. In addition to this, we have to specify how the momentum
components $P^{A}$ commute with the supersymmetry generators. To this end, one
can make reasonable ansaetze for the wanted commutation relations. First of
all, they are restricted by the requirement that the commutation relations
have to be covariant with respect to the action of
$U_{q}(su(2))$. Moreover,
the commutation relations should define an ideal of the algebra
generated by $Q^{\alpha}$, $\bar{Q}^{\alpha}$, $P^{A}$, and the generators of
$U_{q}(su(2))$. In other words, multiplying a relation by a generator and
commuting this generator to the other side of the relation must not change the
relation. This kind of consistency condition completely determines the
commutation relations between the $Q^{\alpha}$, $\bar{Q}^{\alpha}$, and
$P^{A}$.

This way, we found for the commutation relations between $Q^{\alpha}$ and
$\bar{Q}^{\alpha}$ that%
\begin{align}
\bar{Q}^{1}Q^{1}+Q^{1}\bar{Q}^{1} &  =q^{1/2}\lambda_{+}^{1/2}cP^{-}%
,\nonumber\\
\bar{Q}^{1}Q^{2}+q^{-1}Q^{2}\bar{Q}^{1} &  =-q^{-1}\lambda Q^{1}\bar{Q}%
^{2}+qcP^{3},\nonumber\\
\bar{Q}^{2}Q^{1}+q^{-1}Q^{1}\bar{Q}^{2} &  =cP^{3},\nonumber\\
\bar{Q}^{2}Q^{2}+Q^{2}\bar{Q}^{2} &  =q^{1/2}\lambda_{+}^{1/2}cP^{+}%
,\label{qqquereu3}%
\end{align}
where the constant $c$ remains undetermined. We are free to choose $c=1$. The
commutation relations between $P^{A}$ and $Q^{\alpha}$ take the form
\begin{align}
P^{+}Q^{1} &  =q^{-2}Q^{1}P^{+},\nonumber\\
P^{+}Q^{2} &  =Q^{2}P^{+},\nonumber\\
P^{3}Q^{1} &  =q^{-1}Q^{1}P^{3},\nonumber\\
P^{3}Q^{2} &  =q^{-1}Q^{2}P^{3}+q^{-3/2}\lambda\lambda_{+}^{1/2}Q^{1}%
P^{+},\nonumber\\
P^{-}Q^{1} &  =Q^{1}P^{-},\nonumber\\
P^{-}Q^{2} &  =q^{-2}Q^{2}P^{-}+q^{-3/2}\lambda\lambda_{+}^{1/2}Q^{1}%
P^{3}.\label{pqeu3}%
\end{align}
Likewise, we have%
\begin{align}
P^{-}\bar{Q}^{1} &  =\bar{Q}^{1}P^{-},\nonumber\\
P^{-}\bar{Q}^{2} &  =q^{2}\bar{Q}^{2}P^{-},\nonumber\\
P^{3}\bar{Q}^{1} &  =q\bar{Q}^{1}P^{3}-q^{3/2}\lambda\lambda_{+}^{1/2}\bar
{Q}^{2}P^{-},\nonumber\\
P^{3}\bar{Q}^{2} &  =q\bar{Q}^{2}P^{3},\nonumber\\
P^{+}\bar{Q}^{1} &  =q^{2}\bar{Q}^{1}P^{+}-q^{3/2}\lambda\lambda_{+}^{1/2}%
\bar{Q}^{2}P^{3},\nonumber\\
P^{+}\bar{Q}^{2} &  =\bar{Q}^{2}P^{+}.\label{pqquereu3}%
\end{align}

Again, the relations in (\ref{qqquereu3})-(\ref{pqquereu3}) can be written
more compactly by means of $q$-commutators and $q$-anticommutators. In the
case of (\ref{qqquereu3}), for example, one checks that%
\begin{equation}
{\{}\bar{Q}^{\alpha},Q^{\beta}{\}}_{q^{-1}}=\bar{Q}^{\alpha}Q^{\beta}%
+q^{-1}\hat{R}^{\alpha\beta}{}_{\alpha^{\prime}\beta^{\prime}}\,Q^{\alpha
^{\prime}}\bar{Q}^{\beta^{\prime}}=c\,q^{3/2}\lambda_{+}^{-1/2}({\sigma}%
_{A}^{-1})^{\alpha{\beta}}P^{A}.
\end{equation}
To find an analogous form for the relations in (\ref{pqeu3}) and
(\ref{pqquereu3}) we define (cf. Ref.\cite{Maj94})
\begin{equation}
{[}P^{\mu},V{]}_{q}\equiv P_{(1)}^{\mu}VS(P_{(2)}^{\mu}),\qquad{[}P^{\mu}%
,V{]}_{\bar{q}}\equiv P_{(\bar{1})}^{\mu}V\bar{S}(P_{(\bar{2})}^{\mu
}).\label{qComP}%
\end{equation}
The calculation of these $q$-commutators requires to know the Hopf structures
for the momentum algebra. The corresponding coproducts, antipodes, and counits
on the momentum generators read as%
\begin{align}
\Delta(P^{A}) &  =P_{(1)}^{A}\otimes P_{(2)}^{A}=P^{A}\otimes1+\mathcal{L}%
^{A}{}_{B}\otimes P^{B}\nonumber\\
\bar{\Delta}(P^{A}) &  =P_{(\bar{1})}^{A}\otimes P_{(\bar{2})}^{A}%
=P^{A}\otimes1+\mathcal{\bar{L}}^{A}{}_{B}\otimes P^{B},\nonumber\\[0.08in]
S(P^{A}) &  =-S(\mathcal{L}^{A}{}_{B})\,P^{B},\nonumber\\
\bar{S}(P^{A}) &  =-S(\mathcal{\bar{L}}^{A}{}_{B})\,P^{B},\nonumber\\[0.08in]
\epsilon(P^{A}) &  =\bar{\epsilon}(P^{A})=0.\label{HopfCliff}%
\end{align}
Notice that $\mathcal{L}$ and $\mathcal{\bar{L}}$ respectively stand for the
$L$-matrix and its conjugate. Their explicit form can be found in Refs.
\cite{MSW04, BW01}. It should also be mentioned that these $L$-matrices
realize the generators of the quantum group $SU_{q}(2)$ within the algebra
$U_{q}(su(2))$. On these grounds, they depend on generators of $U_{q}(su(2))$
and a unitary scaling operator $\Lambda$ subject to%
\begin{equation}
\Lambda Q^{\alpha}=q^{2}Q^{\alpha}\Lambda,\qquad\Lambda\bar{Q}^{\alpha}%
=q^{2}\bar{Q}^{\alpha}\Lambda.
\end{equation}

Let us now collect all relations of the three-dimensional $q$-deformed
Euclidean superalgebra:%
\begin{gather}
{[}L^{A},L^{B}{]}_{q}=q^{2}\varepsilon^{ABC}g_{CD}\,L^{D},\nonumber\\
{[}L^{A},P^{B}{]}_{q}=q^{2}\varepsilon^{ABC}g_{CD}\,P^{D},\nonumber\\
(P_{A})^{AB}{}_{CD}\,P^{C}P^{D}=0,\nonumber\\
{[}L^{A},Q^{\alpha}{]}_{q}=-q^{-1}\lambda_{+}^{-1/2}(\sigma^{A})_{\beta}%
{}^{\alpha}Q^{\beta},\nonumber\\
{[}L^{A},\bar{Q}^{\alpha}{]}_{q}=-q^{-1}\lambda_{+}^{-1/2}(\sigma^{A})_{\beta
}{}^{\alpha}\bar{Q}^{\beta},\nonumber\\
{[}P^{A},Q^{\alpha}{]}_{\bar{q}}=0,\qquad{[}P^{A},\bar{Q}^{\alpha}{]}%
_{q}=0,\nonumber\\
{\{}Q^{\alpha},Q^{\beta}{\}}_{q}=0,\qquad{\{}\bar{Q}^{\alpha},\bar{Q}^{\beta
}{\}}_{q}=0,\nonumber\\
{\{}\bar{Q}^{\alpha},Q^{\beta}{\}}_{q^{-1}}=q^{3/2}\lambda_{+}^{1/2}({\sigma
}_{A}^{-1})^{\alpha{\beta}}P^{A}.\label{superalgebraeu3}%
\end{gather}

We would like to end this section with some remarks about this algebra. First
of all, the reader should be aware of the fact that $U_{q}(su(2))$ cannot be
generated by $L^{+}$, $L^{3}$, and $L^{-}$ alone. For this reason we have to
add the grouplike operator $\tau^{3}$ and take attention of its commutation
relations with the generators $Q^{\alpha}$, $\bar{Q}^{\alpha}$, and $P^{A}$:%
\begin{align}
\tau^{3}P^{\pm} &  =q^{\mp4}P^{\pm}\tau^{3}, & \tau^{3}P^{3} &  =P^{3}\tau
^{3},\nonumber\\[0.03in]
\tau^{3}Q^{1} &  =q^{2}Q^{1}\tau^{3}, & \tau^{3}Q^{2} &  =q^{-2}Q^{2}\tau
^{3},\nonumber\\
\tau^{3}\bar Q^{1} &  =q^{2}\bar Q^{1}\tau^{3}, & \tau^{3}\bar Q^{2} &
=q^{-2}\bar Q^{2}\tau^{3}.
\end{align}

In the form of (\ref{superalgebraeu3}) the $q$-deformed superalgebra is
strongly reminiscent of its classical counterpart, to which it tends if
$q\rightarrow1$. Notice that in the undeformed limit the $q$-commutators and
$q$-anticommutators then pass into ordinary commutators and anticommutators, respectively.

It should also be mentioned that the algebra in (\ref{superalgebraeu3}) is
compatible with the conjugation assignment%
\begin{gather}
\overline{L^{A}}=g_{AB}L^{B},\qquad\overline{\tau^{3}}=\tau^{3},\qquad
\overline{P^{A}}=g_{AB}P^{B},\nonumber\\
\overline{Q^{\alpha}}=-\varepsilon_{\alpha\beta}\,\bar{Q}^{\beta}%
,\qquad\overline{\bar{Q}^{\alpha}}=\varepsilon_{\alpha\beta}\,Q^{\beta
}.\label{ConSup3}%
\end{gather}
Indeed, one can check that conjugating the relations of our superalgebra and
applying (\ref{ConSup3}) does not change the relations in
(\ref{superalgebraeu3}). In this sense, the three-dimensional $q$-deformed
Euclidean superalgebra is real.

From Ref. \cite{qspinor1} we know of different types of $q$-deformed Pauli and
spin matrices. In this section we used matrices for symmetrized spinors, but
nothing prevents us to work with matrices for antisymmetrized spinors. For the
details we refer the reader to Ref. \cite{qspinor1}.

\section{Symmetry algebras for four-dimensional $q$-deformed Euclidean space}

The considerations for the three-dimensional $q$-deformed Euclidean space carry
over to the four-dimensional one. For this reason, we restrict ourselves
to stating the results, only.

\subsection{The four-dimensional $q$-deformed Euclidean algebra}

The symmetry of four-dimensional $q$-deformed Euclidean space is described by
the quantum algebra $U_{q}(so(4))$. This algebra can be viewed as tensor
product of two commuting copies of $U_{q}(su(2))$, i.e.
\begin{equation}
U_{q}(so(4))\cong U_{q}(su(2))\otimes U_{q}(su(2)).
\end{equation}
The two sets of $U_{q}(su(2))$-generators are denoted by $L_{i}^{\pm}$,
$K_{i}$, $i=1$,$2$, where the $L_{i}^{\pm}$ play the role of $T^{\pm}$ and the
$K_{i}$ that of $(\tau^{3})^{-1/2}$. Thus, the commutation relations between
generators with the same lower index read (cf. Ref. \cite{Oca96})
\begin{gather}
q^{-1}L_{i}^{+}L_{i}^{-}-qL_{i}^{-}L_{i}^{+}=\lambda^{-1}(1-K_{i}^{-2}),\\
L_{i}^{\pm}K_{i}=q^{\mp2}K_{i}L_{i}^{\pm},\qquad i=1,2,\nonumber
\end{gather}
whereas generators with different lower indices always commute.

The generators of $U_{q}(so(4))$ are related to the components $L^{\mu\nu}%
$, $\mu,\nu=1$,$\ldots$,$4$, of an antisymmetric tensor operator (see, for
example, Ref. \cite{Oca96}).\ In Ref. \cite{qliealg} we saw that the $L^{\mu
\nu}$ give rise to a quantum Lie algebra of $U_{q}(so(4)\dot{)}$. The explicit
form of the corresponding $q$-commutators can just as well be found in the work of Ref.
\cite{qliealg}. Here, we only give the general expression
\begin{equation}
{[}L^{\mu\nu},L^{\rho\sigma}{]}_{q}=-q^{-1}\lambda_{+}^{2}(P_{A})^{\mu\nu}%
{}_{\nu^{\prime}\rho^{\prime\prime}}(P_{A})^{\rho\sigma}{}_{\rho^{\prime
}\sigma^{\prime}}g^{\rho^{\prime\prime}\rho^{\prime}}L^{\nu^{\prime}%
\sigma^{\prime}},
\end{equation}
where $P_{A}$ and $g_{\mu\nu}$ respectively stand for an antisymmetrizer and
the quantum metric related to\ $U_{q}(so(4))$.

The quantum algebra $U_{q}(so(4))$ has two Casimir operators. They are%
\begin{align}
C_{1}\equiv &  \;g_{\mu\nu}g_{\rho\sigma}L^{\mu\rho}L^{\nu\sigma}\nonumber\\
= &  \;2L^{23}L^{23}+\lambda_{+}(L^{12}L^{34}+L^{13}L^{24})\nonumber\\
&  +\;q^{2}\lambda_{+}(L^{24}L^{13}+L^{34}L^{12})+(q^{2}+q^{-2})L^{14}%
L^{14}\nonumber\\
&  -\;\lambda(L^{14}L^{23}+L^{23}L^{14}),\label{Cas4dim1}%
\end{align}
and
\begin{align}
C_{2}\equiv &  \;\varepsilon_{\mu\nu\rho\sigma}L^{\mu\nu}L^{\rho\sigma
}\nonumber\\
= &  \;q^{2}\lambda_{+}^{2}(L^{14}L^{23}+L^{23}L^{14})+q^{2}\lambda_{+}%
^{2}(L^{12}L^{34}-L^{13}L^{24})\nonumber\\
&  +\;q^{4}\lambda_{+}^{2}(L^{34}L^{12}-L^{24}L^{13})-q^{2}\lambda\lambda
_{+}^{2}L^{14}L^{14},\label{Cas4dim2}%
\end{align}
where $\varepsilon_{\mu\nu\rho\sigma}$ denotes the totally antisymmetric
tensor of four-dimensional $q$-deformed Euclidean space.

The Euclidean algebra of four-dimensional $q$-deformed Euclidean space is
again a cross product of the quantum algebra $U_{q}(so(4))$ and the momentum
algebra subject to the relations%
\begin{gather}
P^{1}P^{\mu}=qP^{\mu}P^{1},\qquad P^{\mu}P^{4}=qP^{\mu}P^{4},\qquad
\mu=2,3,\nonumber\\
P^{2}P^{3}=P^{3}P^{2},\qquad P^{4}P^{1}=P^{1}P^{4}+\lambda P^{2}P^{3}.
\end{gather}
For the sake of completeness let us note that the above relations can
alternatively be formulated by means of the antisymmetrizer $P_{A}$ of
four-dimensional $q$-deformed Euclidean space:%
\begin{equation}
(P_{A})^{\mu\nu}{}_{\mu^{\prime}\nu^{\prime}}\,P^{\mu^{\prime}}P^{\nu^{\prime
}}=0.
\end{equation}

The momentum components $P^{\mu}$, $\mu=1$,\ldots,$4$, transform under
$U_{q}(so(4))$ as a vector operator. The commutation relations between the
components of the antisymmetric tensor operator $L^{\mu\nu}$ and the momentum
components $P^{\mu}$ were presented in Ref. \cite{qliealg}. The corresponding
$q$-commutators take the general form%
\begin{equation}
{[}L^{\mu\nu},P^{\rho}{]}_{q}=-q^{-1}\lambda_{+}(P_{A})^{\mu\nu}{}%
_{\nu^{\prime}\rho^{\prime}}\;g^{\rho^{\prime}\rho}P^{\nu^{\prime}}.
\end{equation}

The four-dimensional $q$-deformed Euclidean algebra has two Casimir operators.
It should be clear that one Casimir is given by the mass square%
\begin{equation}
m^{2}\equiv g_{\mu\nu}P^{\mu}P^{\nu}.
\end{equation}
To get the second Casimir operator we are looking for a right-vector operator
whose components $W^{\mu}$, $\mu=1$,\ldots,$4$, commute with momenta. Making
suitable ansaetze for the $W^{\mu}$, $\mu=1$,\ldots,$4$, within the Euclidean
algebra we are finally led to%
\begin{align}
W^{1}= &  \;2\lambda^{-1}(K_{1}^{1/2}K_{2}^{-1/2}-K_{1}^{-1/2}K_{2}%
^{1/2})P^{1}\nonumber\\
&  +\;2q^{-1}(K_{1}^{1/2}K_{2}^{-1/2}P^{2}L_{1}^{-}-K_{1}^{-1/2}K_{2}%
^{1/2}P^{3}L_{2}^{-})\\
& \nonumber\\
W^{2}= &  \;2\lambda^{-1}(K_{1}^{1/2}K_{2}^{1/2}-K_{1}^{-1/2}K_{2}%
^{-1/2})P^{2}\nonumber\\
&  +\;2qK_{1}^{1/2}K_{2}^{1/2}P^{1}L_{1}^{+}+2q^{-1}K_{1}^{1/2}K_{2}%
^{1/2}P^{4}L_{2}^{-}\nonumber\\
&  +\;2\lambda K_{1}^{1/2}K_{2}^{1/2}P^{3}L_{2}^{-}L_{1}^{+},\\
& \nonumber\\
W^{3}= &  \;2\lambda^{-1}(K_{1}^{1/2}K_{2}^{1/2}-K_{1}^{-1/2}K_{2}%
^{-1/2})P^{3}\nonumber\\
&  -\;2qK_{1}^{1/2}K_{2}^{1/2}P^{1}L_{2}^{+}-2q^{-1}K_{1}^{1/2}K_{2}%
^{1/2}P^{4}L_{1}^{-}\nonumber\\
&  -\;2\lambda K_{1}^{1/2}K_{2}^{1/2}P^{2}L_{1}^{-}L_{2}^{+},\\
& \nonumber\\
W^{4}= &  \;2\lambda^{-1}(K_{1}^{-1/2}K_{2}^{1/2}-K_{1}^{1/2}K_{2}%
^{-1/2})P^{4}\nonumber\\
&  +\;2q(K_{1}^{-1/2}K_{2}^{1/2}P^{2}L_{2}^{+}-K_{1}^{1/2}K_{2}^{-1/2}%
P^{3}L_{1}^{+}).
\end{align}
Notice that in the undeformed limit these components pass into the classical
expressions $W^{\mu}=\varepsilon^{\mu\nu\rho\sigma}P_{\nu}L_{\rho\sigma}$.
From the $W^{\mu}$ we can build a non-trivial Casimir operator via
\begin{equation}
W^{2}\equiv g_{\mu\nu}W^{\nu}W^{\mu}.
\end{equation}
Written out it reads
\begin{align}
8^{-1}W^{2}= &  \;\lambda^{-2}\lambda_{+}^{-1}(P^{1}P^{4}+qP^{2}%
P^{3})\nonumber\\
&  -\;\lambda^{-2}P^{2}P^{3}(K_{1}K_{2}+q^{2}K_{1}^{-1}K_{2}^{-1})\nonumber\\
&  -\;q\lambda^{-2}P^{1}P^{4}(K_{1}^{-1}K_{2}+K_{1}K_{2}^{-1}-q^{-1}\lambda
K_{1}K_{2})\nonumber\\
&  +\;\lambda^{-1}P^{1}P^{3}(K_{1}K_{2}L_{1}^{+}-K_1K_2^{-1}L_1^+)\nonumber\\
&  +\;\lambda^{-1}P^{1}P^{2}(K_{1}K_{2}L_{2}^{+}-K_1^{-1}K_2L_2^+)\nonumber\\
&  +\;q\lambda^{-1}(P^{2}P^{4}K_{1}(K_{2}-K_{2}^{-1})L_{1}^{-}+P^{3}%
P^{4}(K_{1}-K_{1}^{-1})K_{2}L_{2}^{-})\nonumber\\
&  -\;(P^{2}P^{3}K_{1}K_{2}^{-1}+q^{-1}P^{1}P^{4}K_{1}K_{2})L_{1}^{+}L_{1}%
^{-}\nonumber\\
&  -\;(P^{2}P^{3}K_{1}^{-1}K_{2}+q^{-1}P^{1}P^{4}K_{1}K_{2})L_{2}^{+}L_{2}%
^{-}\nonumber\\
&  -\;q^{-1}(P^{1})^{2}K_{1}K_{2}L_{1}^{+}L_{2}^{+}+(P^{2})^{2}K_{1}K_{2}%
L_{1}^{-}L_{2}^{+}\nonumber\\
&  +\;(P^{3})^{2}K_{1}K_{2}L_{1}^{+}L_{2}^{-}-q(P^{4})^{2}K_{1}K_{2}L_{1}%
^{-}L_{2}^{-}\nonumber\\
&  -\;q^{-2}\lambda(P^{1}P^{2}K_{1}K_{2}L_{1}^{+}L_{1}^{-}L_{2}^{+}+P^{1}%
P^{3}K_{1}K_{2}L_{1}^{+}L_{2}^{+}L_{2}^{-})\nonumber\\
&  -\;q^{-1}\lambda(P^{2}P^{4}K_{1}K_{2}L_{1}^{-}L_{2}^{+}L_{2}^{-}+P^{3}%
P^{4}K_{1}K_{2}L_{1}^{+}L_{1}^{-}L_{2}^{-})\nonumber\\
&  -\;q^{-2}\lambda^2 P^{2}P^{3}K_{1}K_{2}L_{1}^{+}L_{1}^{-}L_{2}^{+}L_{2}^{-}.
\end{align}
This operator commutes with all elements of $U_{q}(so(4))$, since it is
defined as square of a right-vector. The components $W^{\mu}$
commute with momenta, thus the same holds for $W^{2}$.

\subsection{Symmetry algebra in spinor notation \label{SecSpinNotEuc}}

It is sometimes helpful to introduce a new set of generators for
$U_{q}(so(4))$, replacing the generators $L^{\mu\nu}$ with vector indices by
the generators $M^{\alpha\beta}$ and $\tilde{M}^{\dot{\alpha}\dot{\beta}}$
with spinor indices:%
\begin{gather}
M^{\alpha\beta}=k_{1}(\sigma_{\mu\nu}^{-1})^{\alpha\beta}\,L^{\mu\nu
},\nonumber\\
\tilde{M}^{\dot{\alpha}\dot{\beta}}=k_{2}(\tilde{\sigma}_{\mu\nu}^{-1}%
)^{\dot{\alpha}\dot{\beta}}\,L^{\mu\nu},\\[0.08in]
L^{\mu\nu}=k_{1}^{\prime}(\sigma^{\mu\nu})_{\alpha\beta}\,M^{\alpha\beta
}+k_{2}^{\prime}(\tilde{\sigma}^{\mu\nu})_{\dot{\alpha}\dot{\beta}}\,\tilde
{M}^{\dot{\alpha}\dot{\beta}},
\end{gather}
where the constants have to satisfy the condition
\begin{equation}
k_{1}k_{1}^{\prime}=k_{2}k_{2}^{\prime}=\lambda_{+}^{-1}.
\end{equation}
More explicitly, we have%
\begin{gather}
M^{11}=-k_{1}q^{1/2}\lambda_{+}L^{13},\qquad M^{22}=k_{1}q^{1/2}\lambda
_{+}L^{24},\nonumber\\
M^{12}=qM^{21}=k_{1}q^{1/2}(qL^{14}-L^{23}),\\[0.08in]
\tilde{M}^{11}=-k_{2}q^{1/2}\lambda_{+}L^{12},\qquad\tilde{M}^{22}%
=k_{2}q^{1/2}\lambda_{+}L^{34},\nonumber\\
\tilde{M}^{12}=q\tilde{M}^{21}=k_{2}q^{-1/2}L^{14}+k_{2}q^{1/2}L^{23},
\end{gather}
and%
\begin{gather}
L^{12}=-k_{2}^{\prime}q^{-1/2}\tilde{M}^{11},\qquad L^{13}=-k_{1}^{\prime
}q^{-1/2}M^{11},\nonumber\\
L^{14}=k_{2}^{\prime}q^{-1/2}\tilde{M}^{12}+k_{1}^{\prime}q^{-1/2}%
M^{12},\nonumber\\
L^{23}=k_{2}^{\prime}q^{1/2}\tilde{M}^{12}-k_{1}^{\prime}q^{-3/2}%
M^{12},\nonumber\\
L^{24}=k_{1}^{\prime}q^{-1/2}M^{22},\qquad L^{34}=k_{2}^{\prime}%
q^{-1/2}\tilde{M}^{22}.
\end{gather}
The $q$-commutators of the quantum Lie algebra of $U_{q}(so(4))$ in terms of
$M^{\alpha\beta}$ and $\tilde{M}^{\dot{\alpha}\dot{\beta}}$ are given by%
\begin{gather}
{[}M^{\alpha\beta},M^{\gamma\delta}{]}_{q}=(k_{1}^{\prime})^{-1}%
q^{-1}\lambda_{+}\,S^{\alpha\beta}{}_{\beta^{\prime}\alpha^{\prime}%
}S^{\gamma\delta}{}_{\delta^{\prime}\gamma^{\prime}}\,\varepsilon
^{\alpha^{\prime}\delta^{\prime}}M^{\beta^{\prime}\gamma^{\prime}},\nonumber\\
{[}\tilde{M}^{\dot{\alpha}\dot{\beta}},\tilde{M}^{\dot{\gamma}\dot{\delta}}%
{]}_{q}=(k_{2}^{\prime})^{-1}q^{-1}\lambda_{+}\,S^{\dot{\alpha}%
\dot{\beta}}{}_{\dot{\beta}^{\prime}\dot{\alpha}^{\prime}}S^{\dot{\gamma}%
\dot{\delta}}{}_{\dot{\delta}^{\prime}\dot{\gamma}^{\prime}}\,\varepsilon
^{\dot{\alpha}^{\prime}\dot{\delta}^{\prime}}\tilde{M}^{\dot{\beta}^{\prime
}\dot{\gamma}^{\prime}},\nonumber\\[0.03in]
{[}M^{\alpha\beta},\tilde{M}^{\dot{\gamma}\dot{\delta}}{]}_{q}={[}\tilde
{M}^{\dot{\alpha}\dot{\beta}},M^{\gamma\delta}{]}_{q}=0.
\end{gather}
In spinor notation the two Casimirs of $U_{q}(so(4))$ [cf. Eqs.
(\ref{Cas4dim1}) and (\ref{Cas4dim2})] become%
\begin{align}
C_{1}= &  -k_{1}^{\prime2}\lambda_{+}\,\varepsilon_{\alpha\beta}%
\varepsilon_{\alpha^{\prime}\beta^{\prime}}M^{\alpha^{\prime}\alpha}%
M^{\beta\beta^{\prime}}\nonumber\\
&  -k_{2}^{\prime2}\lambda_{+}\,\varepsilon_{\dot{\alpha}\dot{\beta}%
}\varepsilon_{\dot{\alpha}^{\prime}\dot{\beta}^{\prime}}\tilde{M}^{\dot
{\alpha}^{\prime}\dot{\alpha}}\tilde{M}^{\dot{\beta}\dot{\beta}^{\prime}%
},\\[0.1in]
C_{2}= &  \;k_{1}^{\prime2}q^{2}\lambda_{+}^{2}\,\varepsilon_{\alpha\beta
}\varepsilon_{\alpha^{\prime}\beta^{\prime}}M^{\alpha^{\prime}\alpha}%
M^{\beta\beta^{\prime}}\nonumber\\
&  -k_{2}^{\prime2}q^{2}\lambda_{+}^{2}\,\varepsilon_{\dot{\alpha}\dot{\beta}%
}\varepsilon_{\dot{\alpha}^{\prime}\dot{\beta}^{\prime}}\tilde{M}^{\dot
{\alpha}^{\prime}\dot{\alpha}}\tilde{M}^{\dot{\beta}\dot{\beta}^{\prime}}.
\end{align}
These expressions are linked to the Casimir operators of the two
$U_{q}(su(2))$-subalgebras of $U_{q}(so(4))$:%
\begin{align}
C_{1}+q^{-2}\lambda_+^{-1}C_{2} &  =-2k_{2}^{\prime2}\lambda_{+}\varepsilon
_{\dot{\alpha}\dot{\beta}}\varepsilon_{\dot{\alpha}^{\prime}\dot{\beta
}^{\prime}}\tilde{M}^{\dot{\alpha}^{\prime}\dot{\alpha}}\tilde{M}^{\dot{\beta
}\dot{\beta}^{\prime}},\nonumber\\
C_{1}-q^{-2}\lambda_+^{-1}C_{2} &  =-2k_{1}^{\prime2}\lambda_{+}\varepsilon
_{\alpha\beta}\varepsilon_{\alpha^{\prime}\beta^{\prime}}M^{\alpha^{\prime}
\alpha}M^{\beta\beta^{\prime}}.
\end{align}

\subsection{The four-dimensional $q$-deformed Euclidean superalgebra}

In analogy to the three-dimensional case the four-dimensional $q$-deformed
Euclidean superalgebra is generated by the generators of $U_{q}(so(4))$, the
momentum $P^{\mu}$, and the supersymmetry generators\ $Q^{\alpha}$, $\tilde
{Q}^{\dot{\alpha}}$, $\alpha$,$\dot{\alpha}=1$, $2$. Again, the supersymmetry
generators carry spinor indices. Thus, the $Q^{\alpha}$ as well as the
$\tilde{Q}^{\dot{\alpha}}$ span an antisymmetrized quantum plane, i.e.%
\begin{gather}
Q^{\alpha}Q^{\alpha}=\tilde{Q}^{\dot{\alpha}}\tilde{Q}^{\dot{\alpha}}%
=0,\qquad\alpha,\dot{\alpha}=1,2,\nonumber\\
Q^{1}Q^{2}=-q^{-1}Q^{2}Q^{1},\qquad\tilde{Q}^{1}\tilde{Q}^{2}=-q^{-1}\tilde
{Q}^{2}\tilde{Q}^{1}.
\end{gather}

Furthermore, the supersymmetry generators transform as spinor operators under
$U_{q}(so(4))$. More concretely, they refer to the representations $(1/2,0)$
and $(0,1/2)$. On these grounds, the $q$-commutators between the supersymmetry
generators and the $L^{\mu\nu}$ become
\begin{align}
{[}L^{\mu\nu},Q^{\alpha}{]}_{q} &  =-q^{-1}(\sigma^{\mu\nu})_{\beta}{}%
^{\alpha}\,Q^{\beta},\nonumber\\
{[}L^{\mu\nu},\tilde{Q}^{\dot{\alpha}}{]}_{q} &  =-q^{-1}(\tilde{\sigma}%
^{\mu\nu})_{\dot{\beta}}{}^{\dot{\alpha}}\,\tilde{Q}^{\dot{\beta}},
\end{align}
where $\sigma^{\mu\nu}$ and $\tilde{\sigma}^{\mu\nu}$ denote the
two-dimensional spin matrices to four-di\-mensional $q$-deformed Euclidean
space (see Ref. \cite{qspinor1}). For the explicit form of the above
$q$-commutators we refer the reader to Ref. \cite{qliealg}. In spinor notation
we have 
\begin{gather}
\lbrack M^{\alpha\beta},Q^{\gamma}]_{q}=-k_{1}^{\prime -1}q^{-1}
\,S^{\alpha\beta}{}_{\alpha^{\prime}\beta^{\prime}}\,\varepsilon
^{\beta^{\prime}\gamma}Q^{\alpha^{\prime}},\nonumber\\
\lbrack\tilde{M}^{\dot{\alpha}\dot{\beta}},\tilde{Q}^{\dot{\gamma}}%
]_{q}=-k_{2}^{\prime -1}q^{-1}\,S^{\dot{\alpha}\dot{\beta}}{}_{\dot{\alpha}^{\prime}\dot{\beta
}^{\prime}}\,\varepsilon^{\dot{\beta}^{\prime}%
\dot{\gamma}}\tilde{Q}^{\dot{\alpha}^{\prime}},\nonumber\\
\lbrack M^{\alpha\beta},\tilde{Q}^{\dot{\gamma}}]_{q}=[\tilde{M}^{\dot{\alpha
}\dot{\beta}},Q^{\gamma}]_{q}=0.
\end{gather}

Next, we come to the relations between the generators $Q^{\alpha}$ and
$\tilde{Q}^{\dot{\alpha}}$. These relations are covariant with respect
to $\,U_{q}(so(4))$ if they take the form
\begin{align}
\tilde{Q}^{1}Q^{1}+Q^{1}\tilde{Q}^{1} &  =cP^{1},\nonumber\\
\tilde{Q}^{1}Q^{2}+Q^{2}\tilde{Q}^{1} &  =cP^{2},\nonumber\\
\tilde{Q}^{2}Q^{1}+Q^{1}\tilde{Q}^{2} &  =cP^{3},\nonumber\\
\tilde{Q}^{2}Q^{2}+Q^{2}\tilde{Q}^{2} &  =-cP^{4},
\end{align}
where $c$ again denotes an undetermined constant, which we can set equal to 1.
With the help of the Pauli matrices for $q$-deformed Euclidean space in four
dimensions\ (see Ref. \cite{qspinor1} ) the above relations can be
written as ($c=1$)
\begin{equation}
{\{}\tilde{Q}^{\dot{\alpha}},Q^{\beta}{\}}=\;(\tilde{\sigma}_{\mu}%
^{-1})^{\dot\alpha{\beta}}P^{\mu}.
\end{equation}
Notice that the anticommutator in the last formula is the ordinary one.

Last but not least, we have to specify how the momentum operators commute with
the supersymmetry generators. Again, the commutation relations between
momentum operators and supersymmetry generators are uniquely determined by the
requirement that they have to be consistent with our previous relations. This
way, we have
\begin{align}
P^{1}Q^{1}  &  =Q^{1}P^{1},\qquad P^{2}Q^{2}=Q^{2}P^{2},\nonumber\\
P^{3}Q^{1}  &  =Q^{1}P^{3},\qquad P^{4}Q^{2}=Q^{2}P^{4},\nonumber\\
P^{2}Q^{1}  &  =q^{-1}Q^{1}P^{2},\nonumber\\
P^{4}Q^{1}  &  =q^{-1}Q^{1}P^{4},\nonumber\\
P^{1}Q^{2}  &  =q^{-1}Q^{2}P^{1}+q^{-1}\lambda Q^{1}P^{2},\nonumber\\
P^{3}Q^{2}  &  =q^{-1}Q^{2}P^{3}-q^{-1}\lambda Q^{1}P^{4}, \label{PQEuc4}%
\end{align}
and
\begin{align}
P^{1}\tilde{Q}^{1}  &  =\tilde{Q}^{1}P^{1},\qquad P^{2}\tilde{Q}^{1}=\tilde
{Q}^{1}P^{2},\\
P^{3}\tilde{Q}^{2}  &  =\tilde{Q}^{2}P^{3},\qquad P^{4}\tilde{Q}^{2}=\tilde
{Q}^{2}P^{4},\nonumber\\
P^{1}\tilde{Q}^{2}  &  =q\tilde{Q}^{2}P^{1},\nonumber\\
P^{2}\tilde{Q}^{2}  &  =q\tilde{Q}^{2}P^{2},\nonumber\\
P^{3}\tilde{Q}^{1}  &  =q\tilde{Q}^{1}P^{3}-q\lambda\tilde{Q}^{2}%
P^{1},\nonumber\\
P^{4}\tilde{Q}^{1}  &  =q\tilde{Q}^{1}P^{4}+q\lambda\tilde{Q}^{2}P^{2}.\label{PQEuc4Til}
\end{align}

These relations can again be generated from $q$-commutators:%
\begin{equation}
\lbrack P^{\mu},Q^{\alpha}]_{\bar{q}}=0,\qquad\lbrack P^{\mu},\tilde{Q}%
^{\dot{\alpha}}]_{q}=0.
\end{equation}
Notice that the two types of $q$-commutators are defined as in (\ref{qComP}),
but now we have to use the Hopf structures for the four-dimensional $q$-deformed
Euclidean space (see Ref. \cite{MSW04, BW01}). In this respect, it
should be mentioned that these Hopf structures
depend on a unitary scaling operator with%
\begin{equation}
\Lambda Q^{\alpha}=qQ^{\alpha}\Lambda,\qquad\Lambda\tilde{Q}^{\dot{\alpha}%
}=q\tilde{Q}^{\dot{\alpha}}\Lambda.
\end{equation}

Now, we have everything together to write down the $q$-deformed Euclidean
superalgebra in four dimensions:%
\begin{gather}
{[}L^{\mu\nu},L^{\rho\sigma}{]}_{q}=-q^{-1}\lambda_{+}^{2}\,(P_{A})^{\mu\nu}%
{}_{\nu^{\prime}\rho^{\prime\prime}}(P_{A})^{\rho\sigma}{}_{\rho^{\prime
}\sigma^{\prime}}\,g^{\rho^{\prime\prime}\rho^{\prime}}L^{\nu^{\prime}%
\sigma^{\prime}},\nonumber\\
{[}L^{\mu\nu},P^{\rho}{]}_{q}=-q^{-1}\lambda_{+}\,(P_{A})^{\mu\nu}{}%
_{\nu^{\prime}\rho^{\prime}}\,g^{\rho^{\prime}\rho}P^{\nu^{\prime}%
},\nonumber\\
{[}L^{\mu\nu},Q^{\alpha}{]}_{q}=-q^{-1}(\sigma^{\mu\nu})_{\beta}{}^{\alpha
}Q^{\beta},\nonumber\\
{[}L^{\mu\nu},\tilde{Q}^{\dot{\alpha}}{]}_{q}=-q^{-1}(\tilde{\sigma}^{\mu\nu
})_{\dot{\beta}}{}^{\dot{\alpha}}\tilde{Q}^{\dot{\beta}},\nonumber\\
(P_{A})^{\mu\nu}{}_{\mu^{\prime}\nu^{\prime}}\,P^{\mu^{\prime}}P^{\nu^{\prime
}}=0,\nonumber\\
\lbrack P^{\mu},Q^{\alpha}]_{\bar{q}}=0,\qquad\lbrack P^{\mu},\tilde{Q}%
^{\dot{\alpha}}]_{q}=0,\nonumber\\
{\{}Q^{\alpha},Q^{\beta}{\}}_{q}=0,\qquad{\{}\tilde{Q}^{\dot{\alpha}}%
,\tilde{Q}^{\dot{\beta}}{\}}_{q}=0,\nonumber\\
{\{}\tilde{Q}^{\dot{\alpha}},Q^{\beta}{\}}=(\tilde{\sigma}_{\mu}^{-1}%
)^{\dot{\alpha}\beta}P^{\mu}.
\end{gather}
For the sake of completeness we would like to write down how the
symmetry generators $K_i$ commute with the momenta and the supercharges
\begin{align}
K_1P^{1} &  =q^{-1}P^{1}K_1, & K_2P^{1} &  =q^{-1}P^{1}K_2,\nonumber\\
K_1P^{2} &  =qP^{2}K_1, & K_2P^{2} &  =q^{-1}P^{2}K_2,\nonumber\\
K_1P^{3} &  =q^{-1}P^{3}K_1, & K_2P^{3} &  =qP^{3}K_2,\nonumber\\
K_1P^{4} &  =qP^{4}K_1, & K_2P^{4} &  =qP^{4}K_2,\nonumber\\[0.03in]
K_1Q^{1} &  =qQ^{1}K_1, & K_1Q^{2} &  =q^{-1}Q^{2}K_1,\nonumber\\
K_2\tilde Q^{1} &  =q^{-1}\tilde Q^{1}K_2, & K_2\tilde Q^{2} &
=q\tilde Q^{2}K_2.
\end{align}
The remaining relations are trivial.

It should be remarked that conjugating these relations and making use of the
conjugation properties%
\begin{gather}
\overline{L^{\mu\nu}}=g_{\mu\mu^{\prime}}g_{\nu\nu^{\prime}}L^{\nu^{\prime}%
\mu^{\prime}},\qquad\overline{P^{\mu}}=g_{\mu\mu^{\prime}}P^{\mu^{\prime}},\nonumber\\
\overline{Q^{\alpha}}=\varepsilon_{\alpha\alpha^{\prime}}Q^{\alpha^{\prime}%
},\qquad\overline{\tilde{Q}^{\alpha}}=-\varepsilon_{\alpha\alpha^{\prime}%
}\tilde{Q}^{\alpha^{\prime}},
\end{gather}
we obtain a second four-dimensional $q$-deformed Euclidean superalgebra, which
differs from the above one by the commutation relations between momentum
generators and supergenerators, only. Instead of (\ref{PQEuc4}) and
(\ref{PQEuc4Til}) we would have%
\begin{align}
P^{1}Q^{1}  &  =Q^{1}P^{1},\qquad P^{2}Q^{2}=Q^{2}P^{2},\nonumber\\
P^{3}Q^{1}  &  =Q^{1}P^{3},\qquad P^{4}Q^{2}=Q^{2}P^{4},\nonumber\\
P^{1}Q^{2}  &  =qQ^{2}P^{1},\qquad P^{3}Q^{2}=qQ^{2}P^{3},\nonumber\\
P^{2}Q^{1}  &  =qQ^{1}P^{2}-q\lambda Q^{2}P^{1},\nonumber\\
P^{4}Q^{1}  &  =qQ^{1}P^{4}+q\lambda Q^{2}P^{3}.
\end{align}
and%
\begin{align}
P^{1}\tilde{Q}^{1}  &  =\tilde{Q}^{1}P^{1},\qquad P^{2}\tilde{Q}^{1}=\tilde
{Q}^{1}P^{2},\nonumber\\
P^{3}\tilde{Q}^{2}  &  =\tilde{Q}^{2}P^{3},\qquad P^{4}\tilde{Q}^{2}=\tilde
{Q}^{2}P^{4},\nonumber\\
P^{3}\tilde{Q}^{1}  &  =q^{-1}\tilde{Q}^{1}P^{3},\qquad P^{4}\tilde{Q}%
^{1}=q^{-1}\tilde{Q}^{1}P^{4},\nonumber\\
P^{1}\tilde{Q}^{2}  &  =q^{-1}\tilde{Q}^{2}P^{1}+q^{-1}\lambda\tilde{Q}%
^{1}P^{3},\nonumber\\
P^{2}\tilde{Q}^{2}  &  =q^{-1}\tilde{Q}^{2}P^{2}-q^{-1}\lambda\tilde{Q}%
^{1}P^{4}.
\end{align}
In terms of $q$-commutators these relations become%
\begin{equation}
\lbrack P^{\mu},Q^{\alpha}]_{q}=0,\qquad\lbrack P^{\mu},\tilde{Q}^{\dot
{\alpha}}]_{\bar{q}}=0.
\end{equation}

\section{Symmetry algebras for $q$-deformed Min\-kow\-ski space}

In this section we apply our considerations to $q$-deformed Minkowski space
leading us to a $q$-analog of the Poincar\'{e} superalgebra. From a physical
point of view this case is the most interesting one.

\subsection{The $q$-deformed Poincar\'{e} algebra}

The $q$-deformed Poincar\'{e} algebra is a cross product of the $q$-deformed
Lorentz algebra and the momentum algebra of $q$-deformed Minkowski space. In
our work we often use a formulation of $q$-deformed Lorentz algebra as it was
given in Ref. \cite{OSWZ92}. In Ref. \cite{qliealg}, however, we considered as
Lorentz generators the components of an antisymmetric tensor of second rank.
Its components $V^{\mu\nu}$, $\mu$, $\nu\in{\{}0,+,-,3{\}}$, enable us to
formulate the quantum Lie algebra of $q$-deformed Lorentz algebra as
follows:%
\begin{equation}
{[}V^{\mu\nu},V^{\rho\sigma}{]}_{q}=-q^{-1}\lambda_{+}(P_{A})^{\mu\nu}{}%
_{\nu^{\prime}\rho^{\prime\prime}}(P_{A})^{\rho\sigma}{}_{\rho^{\prime}%
\sigma^{\prime}}\,\eta^{\rho^{\prime\prime}\rho^{\prime}}V^{\nu^{\prime}%
\sigma^{\prime}},
\end{equation}
where $\eta_{\mu\nu}$ and $P_{A}$ respectively stand for\ the quantum metric
and an antisymmetrizer of $q$-deformed Minkowski space. For the explicit form of
these $q$-commutators we again refer to Ref. \cite{qliealg}. From the same
reference we know the two Casimirs of $q$-deformed Lorentz algebra:%
\begin{equation}
C_{1}\equiv\eta_{\mu\nu}\eta_{\rho\sigma}V^{\mu\rho}V^{\nu\sigma},\qquad
C_{2}\equiv\varepsilon_{\mu\nu\rho\sigma}V^{\mu\nu}V^{\rho\sigma}.
\end{equation}

The momentum algebra of $q$-deformed Minkowski space is spanned by the
momentum components $P^{\mu}$, $\mu\in{\{}0,+,-,3{\}}$, subject to
\begin{gather}
P^{\mu}P^{0}=P^{0}P^{\mu},\quad\mu\in{\{}0,+,-,3{\},}\nonumber\\
P^{3}P^{\pm}-q^{\pm2}P^{\pm}P^{3}=-q\lambda P^{0}P^{\pm},\nonumber\\
P^{-}P^{+}-P^{+}P^{-}=\lambda(P^{3}P^{3}-P^{0}P^{3}).
\end{gather}
Let us recall that the $P^{\mu}$, $\mu\in{\{}0,+,-,3{\}}$, behave as a
four-vector operator under $q$-deformed Lorentz transformations. For this
reason, the commutation relations between generators of the $q$-deformed
Lorentz algebra and the corresponding momentum components take the form (see
also Ref. \cite{qliealg})%
\begin{equation}
{[}V^{\mu\nu},P^{\rho}{]}_{q}=-q^{-1}(P_{A})^{\mu\nu}{}_{\nu^{\prime}%
\rho^{\prime}}\,\eta^{\rho^{\prime}\rho}P^{\nu^{\prime}}.
\end{equation}

The Casimir operators of $q$-deformed Poincar\'{e} algebra are found from the
same reasonings already applied to the four-dimensional $q$-deformed Euclidean
algebra. In this manner, one Casimir is given by the mass square%
\begin{equation}
m^{2}\equiv g_{\mu\nu}P^{\mu}P^{\nu},
\end{equation}
and the $q$-analog of the spin Casimir becomes \cite{Bloh01, FBM03}
\begin{equation}
W^{2}\equiv\eta_{\nu\mu}W^{\mu}W^{\nu},
\end{equation}
where we have to take as components of the $q$-deformed
Pauli-Lubanski-vector:
\begin{align}
W^{+}= &  \;q^{2}\lambda^{-1}P^{+}-q^{2}\lambda^{-1}P^{+}(\tau^{3})^{1/2}%
(\tau^{1})^{2}-q^{5}\lambda P^{-}(\tau^{3})^{1/2}(T^{2})^{2}\nonumber\\
&  -\;q^{9/2}\lambda_{+}^{1/2}(\tau^{3})^{1/2}P^{3}T^{2}\tau^{1}%
+q^{5/2}\lambda_{+}^{-1/2}(P^{3}-P^{0})T^{+},\\
& \nonumber\\
W^{3}= &  -\;\lambda^{-1}P^{3}-q^{-1/2}\lambda_{+}^{1/2}P^{-}T^{2}\sigma
^{2}-q^{1/2}\lambda_{+}^{1/2}P^{+}S^{1}\tau^{1}\nonumber\\
&  -\;q\lambda\lambda_{+}P^{3}T^{2}S^{1}-q\lambda\lambda_{+}^{-1}(P^{3}%
-P^{0})(\tau^{3})^{1/2}T^{+}T^{-}\nonumber\\
&  +\;q^{-1/2}\lambda_{+}^{-1/2}P^{-}(\tau^{3})^{-1/2}T^{+}-q^{1/2}\lambda
_{+}^{-1/2}P^{+}(\tau^{3})^{-1/2}T^{-}\nonumber\\
&  +\;\lambda^{-1}\lambda_{+}^{-1}((q^{-1}P^{3}-q^{-1}P^{0})(\tau^{3}%
)^{1/2}+(qP^{3}+q^{-1}P^{0})(\tau^{3})^{-1/2}),\\
& \nonumber\\
W^{0}= &  -\lambda^{-1}P^{0}+q^{-1}\lambda\lambda_{+}^{-1}(P^{0}-P^{3})(\tau
^{3})^{-1/2}T^{+}T^{-}\nonumber\\
&  +\;\lambda^{-1}\lambda_{+}^{-1}(q(P^{0}-P^{3})(\tau^{3})^{1/2}+(q^{-1}%
P^{0}+qP^{3})(\tau^{3})^{-1/2})\nonumber\\
&  +\;q^{-1/2}\lambda_{+}^{-1/2}P^{-}(\tau^{3})^{-1/2}T^{+}-q^{1/2}\lambda
_{+}^{-1/2}P^{+}(\tau^{3})^{-1/2}T^{-},\\
& \nonumber\\
W^{-}= &  \;q^{2}\lambda^{-1}P^{-}-q^{-5/2}\lambda_{+}^{-1/2}(P^{3}%
-P^{0})T^{-}-q\lambda P^{+}(\tau^{3})^{-1/2}(S^{1})^{2}\nonumber\\
&  -q^{-1/2}\lambda_{+}^{1/2}P^{3}(\tau^{3})^{-1/2}S^{1}\sigma^{2}%
-q^{-2}\lambda^{-1}P^{-}(\tau^{3})^{-1/2}(\sigma^{2})^{2}.
\end{align}
Note that $S^{1}$, $T^{2}$, $\tau^{1}$, $\sigma^{2}$ denote generators of the
$q$-deformed Lorentz algebra \cite{OSWZ92}. Written out explicitly, the
$q$-deformed spin Casimir reads
\begin{align}
2^{-1}W^{2}= &  \;\lambda^{-2}(q^{-2}(P^{3})^{2}-(P^{0})^{2}+q^{-1}\lambda
P^{3}P^{0}-\lambda_{+}P^{+}P^{-})\nonumber\\
&  +(P^{+})^{2}(\tau^{3})^{-1/2}\;[q\,T^{-}S^{1}\tau^{1}+q^{2}(S^{1}%
)^{2}]\nonumber\\
&  +\lambda_{+}^{-1}(P^{0})^{2}(\tau^{3})^{-1/2}\;[q^{-1}T^{+}T^{-}%
+\lambda^{-2}(q\,\tau^{3}+q^{-1}1)]\nonumber\\
&  -(P^{-})^{2}(\tau^{3})^{-1/2}\;[q^{-1}T^{+}T^{2}\sigma^{2}-q^{4}\tau
^{3}(T^{2})^{2}]\nonumber\\
&  -(P^{3})^{2}(\tau^{3})^{-1/2}\;[q\,\tau^{3}T^{-}T^{2}\tau^{1}-q^{-1}%
\,T^{+}S^{1}\sigma^{2}\nonumber\\
&  -q^{-1}\lambda^{-1}(\sigma^{2})^{2}-\lambda T^{-}T^{2}\sigma^{2}+(\tau
^{3}+q^{2}1-\lambda^{2}T^{+}T^{-})T^{2}%
S^{1}\nonumber\\
&  +\lambda^{-2}\lambda_{+}^{-1}(q\,1+q^{-1}%
\tau^{3}-q^{-1}\lambda^{2}T^{+}T^{-})]\nonumber\\
&  -q^{-3/2}\lambda^{-1}\lambda_{+}^{-1/2}(P^{+}P^{3}-P^{+}P^{0})(\tau
^{3})^{-1/2}\;[\tau^{3}T^{-}(\tau^{1})^{2}\nonumber\\
&  +q\,\tau^{3}S^{1}\tau^{1}-q^{4}T^{-}-q\lambda^{2}T^{+}T^{-}S^{1}\tau
^{1}-q^{2}\lambda^{2}T^{+}(S^{1})^{2}]\nonumber\\
&  -q^{3/2}\lambda^{-1}\lambda_{+}^{-1/2}P^{+}P^{3}(\tau^{3})^{-1/2}%
\;[S^{1}\tau^{1}-q\,S^{1}\sigma^{2}-q^{2}\lambda^{2}\lambda_+T^{-}T^{2}S^{1}%
]\nonumber\\
&  -q^{3/2}\lambda_{+}^{-1/2}P^{+}P^{0}(\tau^{3})^{-1/2}\;[\lambda_{+}%
S^{1}\sigma^{2}+q\lambda^{2}\lambda_+T^{-}T^{2}S^{1}+q^{-2}\lambda S^1\tau^1]\nonumber\\
&  -P^{+}P^{-}(\tau^{3})^{-1/2}\;[T^{+}S^{1}\tau^{1}-T^{-}T^{2}\sigma
^{2}-\lambda^{-2}(q\,\tau^{3}(\tau^{1})^{2})+q^{-1}(\sigma^{2})^{2}%
)]\nonumber\\
&  -P^{3}P^{0}(\tau^{3})^{-1/2}\;[(1-\tau^{3}%
)T^{2}S^{1}-\lambda\lambda_{+}^{-1}(1-\lambda
^{-2}\tau^{3})\nonumber\\
&  +q^{-1}T^{+}S^{1}\sigma^{2}-q\,\tau^{3}T^{-}T^{2}\tau^{1}+q^{-1}%
\lambda^{-1}(\sigma^{2})^{2}\nonumber\\
&  +\lambda T^{-}T^{2}\sigma^{2}+\lambda^{2}T^{+}T^{-}T^{2}S^{1}+2q^{-1}\lambda_+^{-1}T^+T^-]\nonumber\\
&  -q^{-1/2}\lambda_{+}^{-1/2}(P^{3}P^{-}-P^{0}P^{-})(\tau^{3})^{-1/2}%
\;[\lambda^{-1}T^{+}-\lambda^{-1}T^{+}(\sigma^{2})^{2}\nonumber\\
&  +q\lambda^{-1}\tau^{3}T^{2}\sigma^{2}-q\lambda T^{+}T^{-}T^{2}\sigma
^{2}+q^{4}\lambda\tau^{3}T^{-}(T^{2})^{2}]\nonumber\\
&  -q^{1/2}\lambda_{+}^{1/2}P^{3}P^{-}(\tau^{3})^{-1/2}\;[q^{2}\lambda
^{-1}\lambda_{+}^{-1}T^{2}\sigma^{2}\nonumber\\
&  +\lambda T^{+}T^{2}S^{1}-q^{3}\lambda^{-1}\tau^{3}T^{2}\tau^{1}]\nonumber\\
&  -q^{1/2}\lambda^{-1}\lambda_{+}^{-1/2}(\lambda\lambda_{+}-1)P^{0}P^{-}%
(\tau^{3})^{-1/2}\,T^{2}\sigma^{2}.\label{casimirmink}%
\end{align}

\subsection{Symmetry algebra in spinor notation}

In this section we proceed in very much the same way as was done in Sec.
\ref{SecSpinNotEuc} for the four-dimensional $q$-deformed Euclidean algebra.
First, we introduce Lorentz generators with spinor indices%
\begin{gather}
M^{\alpha\beta}=k_{1}(\sigma_{\mu\nu}^{-1})^{\alpha\beta}\,V^{\mu\nu
},\nonumber\\
\bar{M}^{\dot{\alpha}\dot{\beta}}=k_{2}(\bar{\sigma}_{\mu\nu}^{-1}%
)^{\dot{\alpha}\dot{\beta}}\,V^{\mu\nu},\\[0.08in]
V^{\mu\nu}=k_{1}^{\prime}(\sigma^{\mu\nu})_{\alpha\beta}\,M^{\alpha\beta
}+k_{2}^{\prime}(\bar{\sigma}^{\mu\nu})_{\dot{\alpha}\dot{\beta}}\,\bar
{M}^{\dot{\alpha}\dot{\beta}},
\end{gather}
where the two-dimensional spin matrices of $q$-deformed Minkowski space can be
found in Ref. \cite{qspinor1}. For the above identities to be consistent with
each other we have to require that the constants $k_{i}$, $k_{i}^{\prime}$,
$i=1$,$2$, fulfill%
\begin{equation}
k_{1}k_{1}^{\prime}=k_{2}k_{2}^{\prime}=\lambda_{+}^{-1}.
\end{equation}
If we set $k_{1}=k_{2}$, the Lorentz generators $M^{\alpha\beta}$ and $\bar
{M}^{\dot{\alpha}\dot{\beta}}$ show the conjugation properties%
\begin{equation}
\overline{M^{\alpha\beta}}=\varepsilon_{\dot{\alpha}\dot{\alpha}^{\prime}%
}\varepsilon_{\dot{\beta}\dot{\beta}^{\prime}}\bar{M}^{\dot{\beta}^{\prime
}\dot{\alpha}^{\prime}}\qquad\overline{\bar{M}^{\dot{\alpha}\dot{\beta}}%
}=\varepsilon_{\alpha\alpha^{\prime}}\varepsilon_{\beta\beta^{\prime}}\bar
{M}^{\beta^{\prime}\alpha^{\prime}}.
\end{equation}
More explicitly, we have%
\begin{align}
M^{11} &  =k_{1}\lambda_{+}^{1/2}(V^{3-}-V^{0-}),\nonumber\\
M^{12} &  =qM^{21}=k_{1}(q^{-1/2}V^{+-}+q^{1/2}V^{30}),\nonumber\\
M^{22} &  =k_{1}\lambda_{+}^{1/2}(q^{-2}V^{+3}+V^{+0}).\label{SpinFormLorGen}%
\end{align}
and%
\begin{align}
\bar{M}^{11} &  =-k_{2}\lambda_{+}^{1/2}(V^{0-}+q^{-2}V^{3-}),\nonumber\\
\bar{M}^{12} &  =q\bar{M}^{21}=-k_{2}q^{1/2}(qV^{+-}-V^{30}),\nonumber\\
\bar{M}^{22} &  =-k_{2}\lambda_{+}^{1/2}(V^{+3}-V^{+0}%
).\label{SpinFormLorGenCon}%
\end{align}
Solving these equalities for the independent $V^{\mu\nu}$ yields%
\begin{align}
V^{+3} &  =q\lambda_{+}^{-1/2}(k_{1}^{\prime}M^{22}-k_{2}^{\prime}\bar{M}%
^{22}),\nonumber\\
V^{+0} &  =\lambda_{+}^{-1/2}(k_{1}^{\prime}qM^{22}+k_{2}^{\prime}q^{-1}%
\bar{M}^{22}),\nonumber\\
V^{+-} &  =q^{-1/2}(k_{1}^{\prime}M^{12}-k_{2}^{\prime}\bar{M}^{12}%
),\nonumber\\
V^{30} &  =k_{1}^{\prime}q^{1/2}M^{12}+k_{2}^{\prime}q^{-3/2}\bar{M}%
^{12},\nonumber\\
V^{3-} &  =q\lambda_{+}^{-1/2}(k_{1}^{\prime}M^{11}-k_{2}^{\prime}\bar{M}%
^{11}),\nonumber\\
V^{0-} &  =-\lambda_{+}^{-1/2}(k_{1}^{\prime}q^{-1}M^{11}+k_{2}^{\prime}%
q\bar{M}^{11}).
\end{align}
For the quantum Lie algebra in spinor notation we have%
\begin{gather}
{[}M^{\alpha\beta},M^{\gamma\delta}{]}_{q}=-(k_{1}^{\prime})^{-1}%
\,S^{\alpha\beta}{}_{\beta^{\prime}\alpha^{\prime}%
}S^{\gamma\delta}{}_{\delta^{\prime}\gamma^{\prime}}\,\varepsilon
^{\alpha^{\prime}\delta^{\prime}}M^{\beta^{\prime}\gamma^{\prime}},\nonumber\\
{[}\bar{M}^{\alpha\beta},\bar{M}^{\dot{\gamma}\dot{\delta}}{]}_{q}%
=-(k_{2}^{\prime})^{-1}\,S^{\dot{\alpha}\dot{\beta}%
}{}_{\dot{\beta}^{\prime}\dot{\alpha}^{\prime}}S^{\dot{\gamma}\dot{\delta}}%
{}_{\dot{\delta}^{\prime}\dot{\gamma}^{\prime}}\,\varepsilon^{\dot{\alpha
}^{\prime}\dot{\delta}^{\prime}}\bar{M}^{\dot{\beta}^{\prime}\dot{\gamma
}^{\prime}},\nonumber\\
{[}M^{\alpha\beta},\bar{M}^{\dot{\gamma}\dot{\delta}}{]}_{q}={[}\bar{M}%
^{\dot{\alpha}\dot{\beta}},M^{\gamma\delta}{]}_{q}=0.
\end{gather}
In spinor notation the Casimirs of $q$-deformed Lorentz algebra become%
\begin{align}
C_{1}= &  -k_{1}^{\prime2}\lambda_{+}\,\varepsilon_{\alpha\beta}%
\varepsilon_{\alpha^{\prime}\beta^{\prime}}\,M^{\alpha^{\prime}\alpha}%
M^{\beta\beta^{\prime}}\nonumber\\
&  -k_{2}^{\prime}{}^{2}\lambda_{+}\,\varepsilon_{\dot{\alpha}\dot{\beta}%
}\varepsilon_{\dot{\alpha}^{\prime}\dot{\beta}^{\prime}}\,\bar{M}^{\dot
{\alpha}^{\prime}\dot{\alpha}}\bar{M}^{\dot{\beta}\dot{\beta}^{\prime}%
},\\[0.1in]
C_{2}= &  -k_{1}^{\prime2}(3q^{-4}+1+2q^{-1}\lambda)\,\varepsilon_{\alpha
\beta}\varepsilon_{\alpha^{\prime}\beta^{\prime}}\,M^{\alpha^{\prime}\alpha%
}M^{\beta\beta^{\prime}}\nonumber\\
&  +k_{2}^{\prime2}(3q^{-4}+1+2q^{-1}\lambda)\,\varepsilon_{\dot{\alpha}%
\dot{\beta}}\varepsilon_{\dot{\alpha}^{\prime}\dot{\beta}^{\prime}}\,\bar
{M}^{\dot{\alpha}^{\prime}\dot{\alpha}}\bar{M}^{\dot{\beta}\dot{\beta}%
^{\prime}}.
\end{align}
From these formulae we can read off the Casimirs of the two $U_{q}%
(su(2))$-subalgebras:%
\begin{eqnarray}
\lefteqn{(3q^{-4}+1+2q^{-1}\lambda)\lambda_+^{-1}C_{1}+C_{2}= }\nonumber\\  
& =-2k_{1}^{\prime2}(3q^{-4}+1+2q^{-1}\lambda
)\,\varepsilon_{\alpha\beta}\varepsilon_{\alpha^{\prime}\beta^{\prime}%
}\,M^{\alpha^{\prime}\alpha}M^{\beta\beta^{\prime}},\nonumber\\
\lefteqn{(3q^{-4}+1+2q^{-1}\lambda)\lambda_+^{-1}C_{1}-C_{2}= }\nonumber\\  
& =-2k_{2}^{\prime2}(3q^{-4}+1+2q^{-1}\lambda
)\,\varepsilon_{\dot{\alpha}\dot{\beta}}\varepsilon_{\dot{\alpha}^{\prime}%
\dot{\beta}^{\prime}}\,\bar{M}^{\dot{\alpha}^{\prime}}\dot{\alpha}\bar
{M}^{\dot{\beta}\dot{\beta}^{\prime}}.
\end{eqnarray}

\subsection{The $q$-deformed Poincar\'{e} superalgebra}

Now, we extend the $q$-deformed Poincar\'{e} algebra to the $q$-deformed
Poincar\'{e} superalgebra. To this end, we introduce the supersymmetry
generators $Q^{\alpha}$ and $\bar{Q}^{\dot{\alpha}}$. These generators carry
spinor indices, so they fulfill the well-known quantum plane relations:
\begin{gather}
Q^{\alpha}Q^{\alpha}=\bar{Q}^{\dot{\alpha}}\bar{Q}^{\dot{\alpha}}%
=0,\qquad\alpha,\dot{\alpha}=1,2,\\
Q^{1}Q^{2}=-q^{-1}Q^{2}Q^{1},\qquad\bar{Q}^{1}\bar{Q}^{2}=-q^{-1}\bar{Q}%
^{2}\bar{Q}^{1}.\nonumber
\end{gather}
Furthermore, they transform as spinor operators under $q$-deformed Lorentz
transformations. This observation implies the $q$-commutators%
\begin{align}
{[}V^{\mu\nu},Q^{\alpha}{]}_{q} &  =q^{-1}\lambda_{+}^{-1}(\sigma^{\mu\nu
})_{\beta}{}^{\alpha}Q^{\beta},\nonumber\\
{[}V^{\mu\nu},\bar{Q}^{\dot{\alpha}}{]}_{q} &  =q^{-1}\lambda_{+}^{-1}%
(\bar{\sigma}^{\mu\nu})_{\dot{\beta}}{}^{\dot{\alpha}}\bar{Q}^{\dot{\beta}}.
\end{align}
Using Lorentz generators with spinor indices we alternatively have
\begin{gather}
\lbrack M^{\alpha\beta},Q^{\gamma}]_{q}=-(k_{1}^{\prime})^{-1}q^{-1}\lambda_{+}^{-1}%
S^{\alpha\beta}{}_{\alpha^{\prime}\beta^{\prime}}\,\varepsilon^{\beta^{\prime
}\gamma}Q^{\alpha^{\prime}},\nonumber\\
\lbrack\bar{M}^{\dot{\alpha}\dot{\beta}},\bar{Q}^{\dot{\gamma}}]_{q}%
=-(k_{2}^{\prime})^{-1}q^{-1}\lambda_{+}^{-1}S^{\dot{\alpha}\dot{\beta}}{}_{\dot{\alpha}^{\prime}\dot{\beta}^{\prime}%
}\,\varepsilon^{\dot{\beta}^{\prime}\dot{\gamma}%
}\bar{Q}^{\dot{\alpha}^{\prime}},\nonumber\\
\lbrack M^{\alpha\beta},\bar{Q}^{\dot{\gamma}}]_{q}=[\bar{M}^{\dot{\alpha}%
\dot{\beta}},Q^{\gamma}]_{q}=0.
\end{gather}

Next, we turn to the relations between the two supergenerators $Q^{\alpha}$
and $\bar{Q}^{\dot{\alpha}}$. We found
\begin{align}
\bar{Q}^{1}Q^{1}+Q^{1}\bar{Q}^{1} &  =c\,q^{-1}P^{-},\nonumber\\
\bar{Q}^{1}Q^{2}+q^{-1}Q^{2}\bar{Q}^{1} &  =-q^{-1}\lambda Q^{1}\bar{Q}%
^{2}+c\,q^{-1/2}\lambda_{+}^{-1/2}(P^{3}+q^{-2}P^{0}),\nonumber\\
\bar{Q}^{2}Q^{1}+q^{-1}Q^{1}\bar{Q}^{2} &  =-c\,q^{-3/2}\lambda_{+}^{-1/2}%
(P^{0}-P^{3}),\nonumber\\
\bar{Q}^{2}Q^{2}+Q^{2}\bar{Q}^{2} &  =c\,q^{-1}P^{+},
\end{align}
or, for short,
\begin{equation}
{\{}\bar{Q}^{\dot{\alpha}},Q^{\beta}{\}}_{q^{-1}}=\;c(\bar{\sigma}_{\mu}%
^{-1})^{\dot{\alpha}{\beta}}P^{\mu},\label{QQQRelMin}%
\end{equation}
where the $\sigma_{\mu}^{-1}$ denote Pauli matrices for $q$-deformed Minkowski
space (see Ref. \cite{qspinor1}). The constant $c$ can be set equal to $1$.

The commutation relations between momentum generators and supergenerators
read
\begin{align}
P^{+}Q^{1}  &  =q^{-2}Q^{1}P^{+},\quad P^{+}Q^{2}=Q^{2}P^{+},\nonumber\\
P^{0}Q^{1}  &  =q^{-1}Q^{1}P^{0},\quad P^{0}Q^{2}=q^{-1}Q^{2}P^{0},\nonumber\\
P^{3}Q^{1}  &  =q^{-1}Q^{1}P^{3},\nonumber\\
P^{3}Q^{2}  &  =q^{-1}Q^{2}P^{3}+q^{-3/2}\lambda\lambda_{+}^{1/2}Q^{1}%
P^{+},\nonumber\\
P^{-}Q^{1}  &  =Q^{1}P^{-},\nonumber\\
P^{-}Q^{2}  &  =q^{-2}Q^{2}P^{-}+q^{-3/2}\lambda\lambda_{+}^{1/2}Q^{1}P^{3},
\end{align}
and
\begin{align}
P^{0}\bar{Q}^{1}  &  =q\bar{Q}^{1}P^{0},\quad P^{0}\bar{Q}^{2}=q\bar{Q}%
^{2}P^{0},\nonumber\\
P^{-}\bar{Q}^{1}  &  =\bar{Q}^{1}P^{-},\quad P^{-}\bar{Q}^{2}=q^{2}\bar{Q}%
^{1}P^{-},\nonumber\\
P^{3}\bar{Q}^{1}  &  =q\bar{Q}^{1}P^{3}-q^{3/2}\lambda\lambda_{+}^{1/2}\bar
{Q}^{2}P^{-},\nonumber\\
P^{3}\bar{Q}^{2}  &  =q\bar{Q}^{2}P^{3},\nonumber\\
P^{+}\bar{Q}^{1}  &  =q^{2}\bar{Q}^{1}P^{+}-q^{3/2}\lambda\lambda_{+}%
^{1/2}\bar{Q}^{2}P^{3},\nonumber\\
P^{+}\bar{Q}^{2}  &  =\bar{Q}^{2}P^{+}.
\end{align}

Let us note that in very much the same way as was done in Sec.\thinspace
\ref{SecSupAlg3dim} the above commutation relations between momentum
generators and supergenerators can again be written in terms of $q$%
-commutators. To this end, the scaling operator appearing in the Hopf
structure of the momentum generators has to satisfy%
\begin{equation}
\Lambda Q^{\alpha}=q^{-2}Q^{\alpha}\Lambda,\qquad\Lambda\bar{Q}^{\alpha
}=q^{-2}\bar{Q}^{\alpha}\Lambda.
\end{equation}

Now, we have everything together to write down the $q$-deformed Poincar\'{e}
superalgebra:%
\begin{gather}
{[}V^{\mu\nu},V^{\rho\sigma}{]}_{q}=-q^{-1}\lambda_{+}(P_{A})^{\mu\nu}{}%
_{\nu^{\prime}\rho^{\prime\prime}}(P_{A})^{\rho\sigma}{}_{\rho^{\prime}%
\sigma^{\prime}}\eta^{\rho^{\prime\prime}\rho^{\prime}}V^{\nu^{\prime}%
\sigma^{\prime}},\nonumber\\
{[}V^{\mu\nu},P^{\rho}{]}_{q}=-q^{-1}(P_{A})^{\mu\nu}{}_{\nu^{\prime}%
\rho^{\prime}}\eta^{\rho^{\prime}\rho}P^{\nu^{\prime}},\nonumber\\
{[}V^{\mu\nu},Q^{\alpha}{]}_{q}=q^{-1}\lambda_{+}^{-1}(\sigma^{\mu\nu}%
)_{\beta}{}^{\alpha}Q^{\beta},\nonumber\\
{[}V^{\mu\nu},\bar{Q}^{\dot{\alpha}}{]}_{q}=q^{-1}\lambda_{+}^{-1}(\bar
{\sigma}^{\mu\nu})_{\dot{\beta}}{}^{\dot{\alpha}}\bar{Q}^{\dot{\beta}%
},\nonumber\\
(P_{A})^{\mu\nu}{}_{\mu^{\prime}\nu^{\prime}}\,P^{\mu^{\prime}}P^{\nu^{\prime
}}=0,\nonumber\\
{[}P^{\mu},Q^{\alpha}{]}_{q}=0,\qquad{[}P^{\mu},\bar{Q}^{\dot{\alpha}}%
{]}_{\bar{q}}=0,\nonumber\\
{\{}Q^{\alpha},Q^{\beta}{\}}_{q}=0,\qquad{\{}\bar{Q}^{\dot{\alpha}},\bar
{Q}^{\dot{\beta}}{\}}_{q}=0,\nonumber\\
{\{}\bar{Q}^{\dot{\alpha}},Q^{\beta}{\}}_{q^{-1}}=(\bar{\sigma}_{\mu
}^{-1})^{\dot{\alpha}\beta}P^{\mu}.
\end{gather}
This algebra is invariant under the conjugation
\begin{gather}
\overline{P^{\mu}}=\eta_{\mu\nu}P^{\nu},\qquad\overline{Q^{\alpha}%
}=-\varepsilon_{\dot{\alpha}\dot{\beta}}\bar{Q}^{\dot{\beta}},\qquad
\overline{\bar{Q}^{\dot{\alpha}}}=\varepsilon_{\alpha\beta}Q^{\beta
},\nonumber\\[0.02in]
\overline{V^{\mu\nu}}=(-1)^{\delta_{\mu0}+\delta_{\nu0}}\eta_{\mu\mu^{\prime}%
}\eta_{\nu\nu^{\prime}}V^{\nu^{\prime}\mu^{\prime}}.
\end{gather}

\section{Conclusion}

Let us end with some comments on what we have done so far. We considered the
$q$-deformed Poincar\'{e} algebra and the $q$-deformed Euclidean algebra in
three and four dimensions. These algebras describe the symmetry of
$q$-deformed Minkowski space and the $q$-deformed Euclidean spaces in three
and four dimensions. We extended these algebras to superalgebras by adding two
supersymmetry generators with spinor indices. Exploiting consistency arguments
we could determine all commutation relations concerning the supersymmetry
generators. Furthermore, we were able to write down our $q$-deformed
superalgebras in a way that reveals striking similarities to their undeformed
counterparts. To achieve this we introduced generators with definite
transformation properties and defined their adjoint actions as $q$-commutators.

Lastly, we would like to point out that the $q$-deformed Poincar\'{e}
superalgebra should be useful\ in $q$-deforming supersymmetric models. To this
end, we reconsider Eq. (\ref{QQQRelMin}) and contract it with the Pauli matrix
$(\bar{\sigma}^{\mu})_{\dot{\alpha}\beta}$. In doing so we obtain
\begin{align}
P^{\mu} &  =(\bar{\sigma}^{\mu})_{\dot{\alpha}\beta}\big[\bar{Q}^{\dot{\alpha
}}Q^{\beta}+q^{-1}\hat{R}^{\dot{\alpha}\beta}{}_{\beta^{\prime}\dot{\alpha
}^{\prime}}\,Q^{\beta^{\prime}}\bar{Q}^{\dot{\alpha}^{\prime}}\big]\nonumber\\
&  =\big[(\bar{\sigma}^{\mu})_{\dot{\alpha}\beta}\,\bar{Q}^{\dot{\alpha}%
}Q^{\beta}+q^{-2}(\sigma^{\mu})_{\beta\dot{\alpha}}\,Q^{\beta}\bar{Q}%
^{\dot{\alpha}}\big].\label{Psup}%
\end{align}
Notice that the second equality in (\ref{Psup}) holds due to (see
Ref. \cite{qspinor1})%
\begin{equation}
(\sigma^{\mu})_{\beta^{\prime}\dot{\alpha}^{\prime}}=q(\bar{\sigma}^{\mu
})_{\dot{\alpha}\beta}\,\hat{R}^{\dot{\alpha}\beta}{}_{\beta^{\prime}%
\dot{\alpha}^{\prime}}.
\end{equation}
From (\ref{Psup}) we get the supersymmetric Hamiltonian%
\begin{align}
H\equiv P^{0}= &  \;\big[(\bar{\sigma}^{0})_{\dot{\alpha}\beta}\,\bar{Q}%
^{\dot{\alpha}}Q^{\beta}+q^{-2}(\sigma^{0})_{\beta\dot{\alpha}}\,Q^{\beta}%
\bar{Q}^{\dot{\alpha}}\big]\nonumber\\
= &  \;\big[-q^{-1/2}\,\bar{Q}^{1}Q^{2}+q^{1/2}\,\bar{Q}^{2}Q^{1}\nonumber\\
&  +q^{-5/2}\,Q^{1}\bar{Q}^{2}-q^{-3/2}\,Q^{2}\bar{Q}^{1}\big].
\end{align}
The conjugation properties of $Q^{\alpha}$ and $\bar{Q}^{\dot{\alpha}}$ imply%
\begin{equation}
\overline{H}=\overline{P^{0}}=P^{0}=H,
\end{equation}
i.e. our $q$-deformed supersymmetric Hamiltonian is a real operator. It should
also be mentioned that $H$ does not commute with the supersymmetry generators,
since we have
\begin{gather}
P^{0}Q^{1}=q^{-1}Q^{1}P^{0},\quad P^{0}Q^{2}=q^{-1}Q^{2}P^{0},\nonumber\\
P^{0}\bar{Q}^{1}=q\bar{Q}^{1}P^{0},\quad P^{0}\bar{Q}^{2}=q\bar{Q}^{2}P^{0}.
\end{gather}

\noindent\textbf{Acknowledgements}\newline First of all we are very grateful
to Eberhard Zeidler for his invitation to the MPI Leipzig, his special
interest in our work, and his financial support. Furthermore we would like to
thank Fabian Bachmaier and Ina Stein for their support. Finally, we thank
Dieter L\"{u}st for kind hospitality.

\appendix

\section{Invariant tensors of $q$-deformed quantum spaces}

The aim of this appendix is the following. For the quantum spaces under
consideration we list the non-vanishing \ components of the quantum metric and
the totally antisymmetric tensor.

The non-vanishing elements of the two-dimensional spinor metric have the
values%
\begin{equation}
\varepsilon^{12}=q^{-1/2},\quad\varepsilon^{21}=-q^{1/2}.
\end{equation}
The spinor metric is antisymmetric in a $q$-deformed sense and its inverse is
given by
\begin{equation}
(\varepsilon^{-1})^{ij}=\varepsilon_{ij}=-\varepsilon^{ij}.
\end{equation}

The non-vanishing elements of the three-dimensional Euclidean quantum metric
are
\begin{equation}
g^{+-}=-q,\quad g^{33}=1,\quad g^{-+}=-q^{-1}. \label{metriceu3app}%
\end{equation}
For its inverse $g_{AB}$ we have%
\begin{equation}
g_{AB}=g^{AB}.
\end{equation}
The non-vanishing components of the three-dimensional $q$-deformed epsilon
tensor take the form%
\begin{align}
\varepsilon^{-3+}  &  =-q^{-4}, & \varepsilon^{3-+}  &  =q^{-2},\nonumber\\
\varepsilon^{-+3}  &  =q^{-2}, & \varepsilon^{+-3}  &  =-q^{-2},\nonumber\\
\varepsilon^{3+-}  &  =-q^{-2}, & \varepsilon^{+3-}  &  =1.\nonumber\\
\varepsilon^{333}  &  =-q^{-2}\lambda. &  &  \label{epseu3app}%
\end{align}
The elements of the lower indexed epsilon tensor we get from  the
identification
\begin{align}
\varepsilon_{ABC}=\varepsilon^{CBA}.
\end{align}
Next we come to four-dimensional $q$-deformed Euclidean space. Its metric has
the non-vanishing components
\begin{equation}
g^{14}=q^{-1},\quad g^{23}=g^{32}=1,\quad g^{41}=q.\label{metriceu4app}%
\end{equation}
Its inverse is denoted by $g_{\mu\nu}$ and fulfills
\begin{equation}
g_{\mu\nu}=g^{\mu\nu}.
\end{equation}
The non-vanishing components of the epsilon tensor of four-dimensional
$q$-deformed Euclidean space are%
\begin{align}
\varepsilon^{1234} &  =1, & \varepsilon^{1432} &  =-q^{2}, & \varepsilon
^{2413} &  =-q^{2},\nonumber\\
\varepsilon^{2134} &  =-q, & \varepsilon^{4132} &  =q^{2}, & \varepsilon
^{4213} &  =q^{3},\nonumber\\
\varepsilon^{1324} &  =-1, & \varepsilon^{3412} &  =q^{2}, & \varepsilon
^{2341} &  =-q^{2},\nonumber\\
\varepsilon^{3124} &  =q, & \varepsilon^{4312} &  =-q^{3}, & \varepsilon
_{3241} &  =q^{2},\nonumber\\
\varepsilon^{2314} &  =q^{2}, & \varepsilon^{1243} &  =-q, & \varepsilon
^{2431} &  =q^{3},\nonumber\\
\varepsilon^{3214} &  =-q^{2}, & \varepsilon^{2143} &  =q^{2}, &
\varepsilon^{4231} &  =-q^{4},\nonumber\\
\varepsilon^{1342} &  =q, & \varepsilon^{1423} &  =q^{2}, & \varepsilon^{3421}
&  =-q^{3},\nonumber\\
\varepsilon^{3142} &  =-q^{2}, & \varepsilon^{4123} &  =-q^{2}, &
\varepsilon^{4321} &  =q^{4},\label{epseu4app}%
\end{align}
together with the non-classical components%
\begin{equation}
\varepsilon^{3232}=-\varepsilon^{2323}=-q^{2}\lambda.
\end{equation}

The quantum metric of $q$-deformed Min\-kow\-ski space is given by%
\begin{equation}
\eta^{00}=-1,\quad\eta^{33}=1,\quad\eta^{+-}=-q,\quad\eta^{-+}=-q^{-1}%
,\label{metricminkapp}%
\end{equation}
with inverse%
\begin{equation}
\eta_{\mu\nu}=\eta^{\mu\nu}.
\end{equation}
As non-vanishing components of the corresponding epsilon tensor we have%
\begin{align}
\varepsilon^{+30-} &  =1, & \varepsilon^{+-03} &  =-q^{-2}, & \varepsilon
^{3-+0} &  =q^{-2},\nonumber\\
\varepsilon^{3+0-} &  =-q^{-2}, & \varepsilon^{-+03} &  =q^{-2}, &
\varepsilon^{-3+0} &  =q^{-4},\nonumber\\
\varepsilon^{+03-} &  =-1, & \varepsilon^{0-+3} &  =q^{-2}, & \varepsilon
^{30-+} &  =-q^{-2},\nonumber\\
\varepsilon^{0+3-} &  =1, & \varepsilon^{-0+3} &  =-q^{-2}, & \varepsilon
^{03-+} &  =q^{-2},\nonumber\\
\varepsilon^{30+-} &  =q^{-2}, & \varepsilon^{+3-0} &  =-1, & \varepsilon
^{3-0+} &  =q^{-2},\nonumber\\
\varepsilon^{03+-} &  =-q^{-2}, & \varepsilon^{3+-0} &  =q^{-2}, &
\varepsilon^{-30+} &  =-q^{-4},\nonumber\\
\varepsilon^{+0-3} &  =q^{-2}, & \varepsilon^{+-30} &  =q^{-2}, &
\varepsilon^{0-3+} &  =-q^{-4},\nonumber\\
\varepsilon^{0+-3} &  =-q^{-2}, & \varepsilon^{-+30} &  =-q^{-2}, &
\varepsilon^{-03+} &  =q^{-4},\label{epsminkapp}%
\end{align}
and%
\begin{align}
\varepsilon^{0-0+} &  =q^{-3}\lambda, & \varepsilon^{-0+0} &  =-q^{-3}%
\lambda,\nonumber\\
\varepsilon^{0333} &  =-q^{-2}\lambda, & \varepsilon^{3330} &  =q^{-2}%
\lambda,\nonumber\\
\varepsilon^{3033} &  =+q^{-2}\lambda, & \varepsilon^{3030} &  =-q^{-2}%
\lambda,\nonumber\\
\varepsilon^{3303} &  =-q^{-2}\lambda, & \varepsilon^{+0-0} &  =-q^{-1}%
\lambda,\nonumber\\
\varepsilon^{0303} &  =q^{-2}\lambda, & \varepsilon^{0+0-} &  =q^{-1}\lambda.
\end{align}

Lowering the indices of the epsilon tensor is achieved by the quantum metric.
In this manner we have, for example,%
\begin{equation}
\varepsilon_{\mu\nu\rho\sigma}=\eta_{\mu\mu^{\prime}}\eta_{\nu\nu^{\prime}%
}\eta_{\rho\rho^{\prime}}\eta_{\sigma\sigma^{\prime}}\,\varepsilon
^{\mu^{\prime}\nu^{\prime}\rho^{\prime}\sigma^{\prime}}.
\end{equation}

\end{document}